\documentclass[sigconf, nonacm]{acmart}

\AtBeginDocument{%
  }

\usepackage{latexsym}
\usepackage{subcaption}
\usepackage{multirow}
\usepackage{enumitem}
\usepackage{pifont}
\usepackage{makecell}
\usepackage{colortbl}
\usepackage{arydshln}
\usepackage{marvosym}

\definecolor{rvBlue1}{HTML}{1A5A8A}
\definecolor{rvBlue2}{HTML}{3D88B5}
\definecolor{rvBlue3}{HTML}{5FB1D3}
\definecolor{rvOrange1}{HTML}{C57220}
\definecolor{rvOrange2}{HTML}{D58836}
\definecolor{rvOrange3}{HTML}{E59E4D}
\definecolor{rvOrange4}{HTML}{ECAF5E}
\definecolor{rvOrange5}{HTML}{F4C075}
\newcommand{\ours}{\texttt{RecVerse}}

\newcommand{\eg}{\emph{e.g.}}
\newcommand{\ie}{\emph{i.e.}}

\begin{document}

\title{Towards Faithful Simulation of Human Shopping Behavior}

\author{\mbox{Jiakai Tang\textsuperscript{1,4,$*$,$\dag$}}, \mbox{Yan Mi\textsuperscript{4,$*$}}, \mbox{Jing Yu\textsuperscript{4,$*$}}, \mbox{Yang Zhang\textsuperscript{3}}, \mbox{See-Kiong Ng\textsuperscript{3}}, \mbox{Qi Cao\textsuperscript{2}}, \mbox{Fei Sun\textsuperscript{2,\Letter}}, \mbox{Xu Chen\textsuperscript{1,\Letter}}, \mbox{Wen Chen\textsuperscript{4,\Letter}}, \mbox{Jian Wu\textsuperscript{4}}, \mbox{Han Zhu\textsuperscript{4}}, \mbox{Bo Zheng\textsuperscript{4}}}
\affiliation{%
  \institution{\textsuperscript{1}Gaoling School of Artificial Intelligence, Renmin University of China, Beijing, China\\
    \textsuperscript{2}University of Chinese Academy of Sciences, Beijing, China\\
    \textsuperscript{3}National University of Singapore, Singapore\\
    \textsuperscript{4}Alibaba Group, Beijing, China}
  \city{}
  \country{}
}
\email{tangjiakai5704@ruc.edu.cn}
\renewcommand{\shortauthors}{Jiakai Tang et al.}

\begin{abstract}
Simulating realistic user shopping behavior underpins offline evaluation and reinforcement learning in e-commerce scenarios.
While recent LLM- and VLM-based simulators have made encouraging progress, reproducing a real browsing session remains difficult for two reasons.
(i) \textbf{Memory Challenge}: a shopping session spans dozens of pages, yet existing agents either discard long-range observation histories, losing the evolving user state, or naively concatenate them, overwhelming the context window and even degrading simulation quality.
(ii) \textbf{Optimization Challenge}: current user simulators are typically supervised to match each logged action via imitation or step-level rewards; the resulting sessions often display unrealistic patterns, such as over-exploration or excessive passivity, which per-step supervision can neither detect nor correct.

To address the above challenges, we present \textbf{\textit{\ours{}}}, a GUI-grounded simulation agent that perceives pages through screenshots and produces faithful multi-turn trajectories.
\textbf{For the memory challenge}, \ours{} adopts a cognitive-inspired hierarchical memory: \emph{Working Memory} for short-term focus, \emph{Episodic Memory} for in-session traces, and \emph{Preference Memory} for high-level intent, with memory updates treated as actions so that the agent adaptively learns \emph{when} and \emph{what} to memorize.
\textbf{For the optimization challenge}, \ours{} is optimized with a trajectory-level RL objective that scores entire sessions, aligning both macro-level action-type distributions and micro-level shopping intent with real users.
We further release \textbf{USB} (\textbf{U}ser \textbf{S}imulation \textbf{B}enchmark), an interactive e-commerce GUI trajectory dataset for multi-turn user simulation.
Experiments show that \ours{} significantly outperforms existing baselines in both behavioral fidelity and intent consistency.
\end{abstract}



\maketitle

\hypersetup{
  pdfauthor={Jiakai Tang, Yan Mi, Jing Yu, Yang Zhang, See-Kiong Ng, Qi Cao, Fei Sun, Xu Chen, Wen Chen, Jian Wu, Han Zhu, Bo Zheng},
  pdftitle={Towards Faithful Simulation of Human Shopping Behavior}
}

{\renewcommand{\thefootnote}{$*$}\footnotetext{Co-first authors.}}
{\renewcommand{\thefootnote}{$\ddagger$}\footnotetext{Project Lead.}}
{\renewcommand{\thefootnote}{\Letter}\footnotetext{Corresponding authors.}}

\section{Introduction}
\label{sec:intro}

User behavior simulation, the task of generating realistic browsing trajectories that reflect how humans explore, compare, and purchase products, is a longstanding research topic in e-commerce recommendation~\citep{ie2019recsim,shi2019virtual}.
By acting as a controllable proxy for real shoppers, faithful simulators support offline and counterfactual policy evaluation that would otherwise require costly online experiments~\citep{mladenov2021recsimng,joachims2017unbiased,saito2021open}, and serve as interactive environments for RL-based recommendation training.
As recommender systems grow more interactive and personalized, and as the rapid progress of large language models (LLMs)~\cite{10.1145/3758091,shao2023character,piao2025agentsociety} and agent techniques~\cite{tang-etal-2025-gensim,tang2025interactive,yao2022webshop} opens new avenues for behavior modeling, building simulators that faithfully approximate human shopping behavior has become both an increasingly pressing need and a newly tractable opportunity~\cite{guozhen2024human,mou2026individual,wang2024survey,gao2024large}.

A growing body of work has pursued this goal through three main lines, each lifting the simulator's observation modality closer to what real users perceive, yet each falling short.
\textbf{(i) Rule-based simulators}~\citep{ie2019recsim,mladenov2021recsimng,shi2019virtual} construct probabilistic user models over vectorized item features and user states, but their pre-specified action spaces cannot represent the complexity of modern e-commerce interfaces.
\textbf{(ii) LLM-based agents}~\citep{wang2025user,zhang2024generative,zhang2024agentcf} bring reasoning and persona modeling to the task, yet they still operate on text descriptions or metadata, ignoring the image-rich, spatially structured pages that drive real browsing decisions.
\textbf{(iii) GUI-grounded simulators}~\citep{Zhang2025SeeTA,Zhang2026ExploringRS} finally close the modality gap by consuming page screenshots, but they either depend on prompt engineering without task-specific adaptation~\citep{Zhang2026ExploringRS}, or condition on the current screenshot with heuristically pruned textual histories and step-level optimization~\citep{Zhang2025SeeTA}.
Crucially, all three lines share a deeper gap beyond their individual flaws: none is built to handle long-term simulation.

\begin{figure}[t]
  \centering
  \includegraphics[width=0.9\columnwidth]{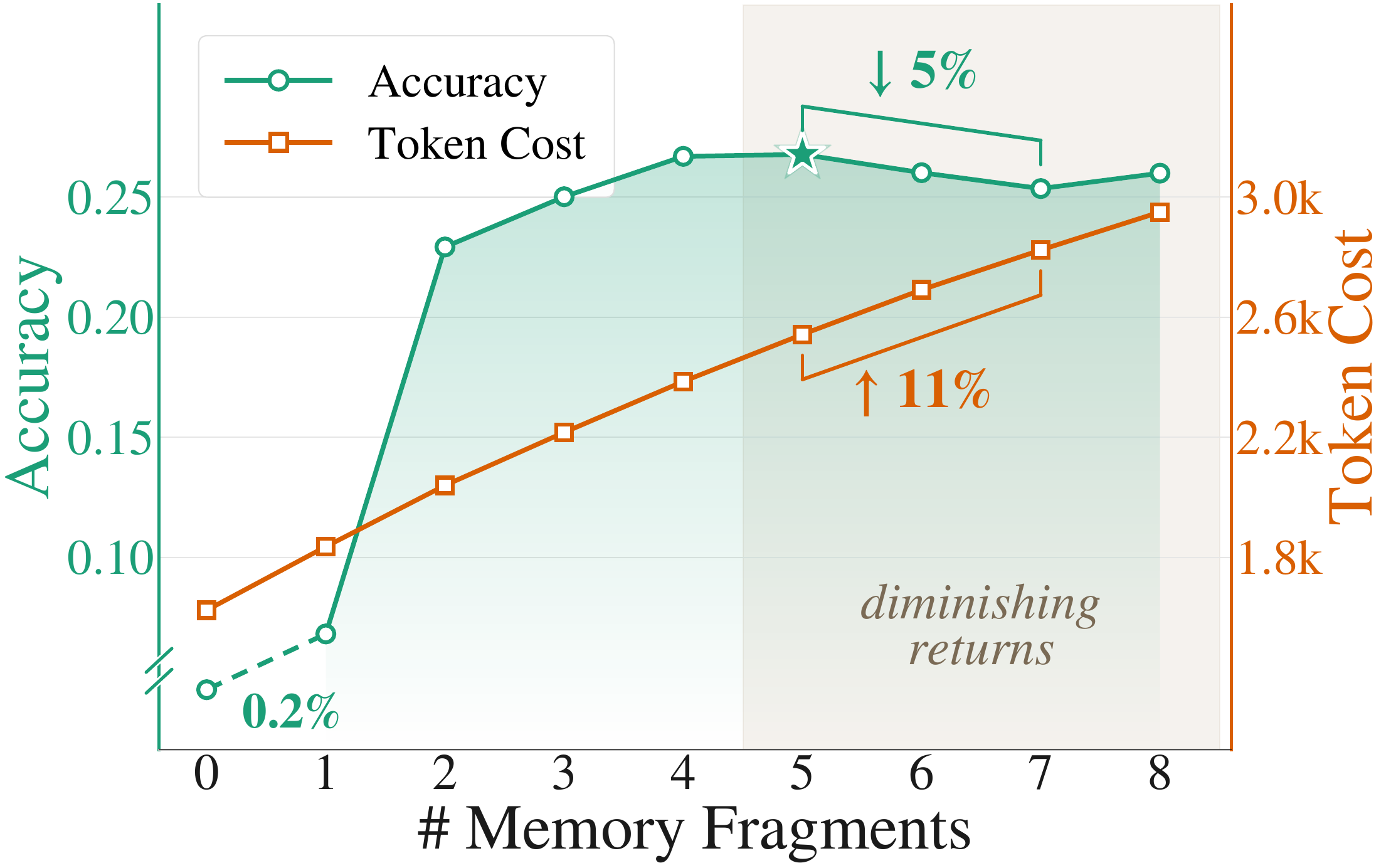}
  \vspace{-10pt}
  \caption{Effect of increasing visual memory fragments in the multimodal context. Accuracy improves initially but exhibits diminishing returns and slight degradation beyond a moderate context length, while token cost continues to grow.}
  \label{fig:context_degradation}
  \vspace{-15pt}
\end{figure}

Building a faithful user simulator, however, requires solving two key problems that are not well-addressed by existing work:

$\diamond$ \textbf{The memory problem.}
A typical shopping session spans dozens of viewport frames, and a click or purchase often traces back to items browsed and compared many pages earlier.
Existing approaches either discard or heuristically prune history~\citep{Zhang2025SeeTA}, or na\"ively concatenate it into the context~\citep{yao2023react}; the former severs these cross-page dependencies, while the latter overwhelms the context window~\citep{openai2023gpt4,qwen2024qwen2vl} and, as long-context studies~\citep{liu2024lostmiddle,wang2023longmem} and our experiments (Figure~\ref{fig:context_degradation}) show, \emph{degrades} simulation quality as context grows.
Cognitive science suggests a principled alternative: humans rely on complementary memory systems~\citep{atkinson1968human,tulving1985many,baddeley2000episodic}, from fleeting visual impressions to session-level behavioral tracking to stable personal preferences.
Yet which impressions will matter later is latent and not annotated, and the heuristic memory management of prior memory-augmented agents~\citep{shinn2023reflexion,zhong2024memorybank,wang2024voyager} cannot decide \emph{when} and \emph{what} to remember during continuous browsing.

$\diamond$ \textbf{The optimization problem.}
User simulation methods universally adopt next-action imitation~\citep{wang2025user,Zhang2025SeeTA}: predicting the next action given the current state.
This paradigm is misaligned with behavioral fidelity:
\textbf{(a)} real browsing contains substantial \emph{behavior noise} (\eg, accidental over-scrolls followed by immediate reversal), and fitting such noise confuses the model's reasoning about user intent;
\textbf{(b)} step-wise supervision offers no incentive for realistic overall behavior: the resulting simulator often deviates from real users in how frequently it clicks and explores across a session.
Faithful simulation requires alignment at two granularities: \textbf{\emph{macro-level}}, where the agent's overall behavioral patterns match real user distributions (\eg, interaction frequency); and \textbf{\emph{micro-level}}, where the agent's interactions align with real user intent (\eg, product category).

We present \textbf{\textit{\ours{}}}, a user behavior simulation agent that addresses both problems above.
Building on the GUI-grounded direction, \ours{} perceives recommendation pages through pixel-level screenshots, aligning its observation modality with what real users actually see.
\textbf{For the memory problem}, we introduce a cognitive-inspired multi-level memory system that progressively compresses browsing history, from recent frames in \emph{Working Memory}, to textual interaction records in \emph{Episodic Memory}, and finally to distilled user intent in \emph{Preference Memory}, so that long-range dependencies are preserved in increasingly abstract forms while the context stays bounded.
We further cast memory updates as actions and optimize them jointly with browsing behavior via RL, so that the agent itself learns \emph{when} and \emph{what} to remember.
\textbf{For the optimization problem}, a trajectory-level RL framework moves beyond step-wise behavior cloning with two complementary rewards: a \emph{distribution-aligned reward} that prevents the agent from deviating from real user behavioral patterns at the macro level, and an \emph{intent-aware reward} that enforces micro-level consistency on the agent's interaction decisions.
We also release \textbf{USB}, a GUI-grounded e-commerce benchmark with \textbf{5,274} real user trajectories pairing page-level screenshots; unlike prior datasets restricted to static offline logs, it offers, to our knowledge, the first real-world interactive environment supporting multi-turn agentic reinforcement learning.

In summary, our contributions are:
\begin{itemize}[leftmargin=1em]
    \item We formalize GUI-grounded user behavior simulation and release a large-scale benchmark with pixel-level observations and action annotations designed for long-horizon reinforcement learning.
    \item We propose \ours{}, integrating a cognitive-inspired multi-level memory system with learned operations and a trajectory-focused RL framework that aligns simulated behavior with real users at both macro and micro levels.
    \item We conduct extensive experiments on real e-commerce sessions, showing that \ours{} significantly outperforms existing baselines in behavioral fidelity and intent consistency.
\end{itemize}

\section{Preliminary}
\label{sec:preliminary}

\subsection{Task Definition}
\label{sec:task}

In this paper, we study user behavior simulation for e-commerce scenario.
Given a real user session trajectory
\begin{equation}
    \tau = \{(o_1, a_1), (o_2, a_2), \ldots, (o_T, a_T)\},
\end{equation}
where $o_t$ denotes the page observation (\eg, item descriptions, contextual information, visual screenshots) and $a_t \in \mathcal{A}$ is a user action (\eg, scroll, click, and add-to-cart), the goal is to train an agent $\pi_\theta$ that emulates the user's browsing trajectory $\tau$ by producing actions faithful to real user browsing behavior, conditioned on the current observation $o_t$ and the historical context $h_{<t} = \{(o_i, a_i)\}_{i=1}^{t-1}$.

In this work, we focus on a GUI-grounded setting where $o_t$ takes the form of page-level screenshots, grounding the agent in the same visual modality through which real users perceive recommendation interfaces.
This setting differs fundamentally from task-oriented GUI agents~\citep{hong2024cogagent,cheng2024seeclick,zheng2024gpt4vision,bai2024digirl} that are \textcolor{green!70!black}{\emph{goal-oriented}} (\eg, ``find and buy item X'').
User simulation is instead \textcolor{red!70!black}{\emph{process-oriented}}: it aims to capture how users explore, hesitate, compare, and eventually act.
This shift manifests along three axes:
\begin{itemize}[leftmargin=1.2em, itemsep=2pt, topsep=2pt, label=$\diamond$]
    \item \textbf{Off-task behavior as signal.} Behaviors that goal-oriented agents suppress as ``failure modes'' (\eg, repeated comparison) constitute the very signal a faithful simulator aims to reproduce.
    \item \textbf{Trajectory distribution, not single optimum.} Rather than converging to a single optimal trajectory, the simulator targets the multi-modal distribution of how heterogeneous users navigate similar or different intents over the same interface.
    \item \textbf{Cognitive state, not explicit subgoals.} Instead of tracking discrete subgoals toward a known endpoint, the simulator reasons over long-horizon cognitive states that shape users' mindsets.
\end{itemize}
Therefore, these shifts move both supervision and evaluation from end-task success toward trajectory-level distributional fidelity and fine-grained intent alignment with real shoppers.

\subsection{Imitation Learning}
\label{sec:sft}

Imitation Learning (IL) trains an agent to reproduce expert behavior from pre-collected demonstrations~\citep{lu2025can,bougie2026alignuser}.
In user simulation, this amounts to maximizing the log-likelihood of real user actions given the observation and history:
\begin{equation}
    \mathcal{L}_{\text{IL}} = -\sum_{t=1}^{T} \log \pi_\theta(a_t \mid o_t, h_{<t})
\end{equation}
where $h_{<t}$ denotes the historical context available at step $t$.
IL provides a straightforward behavioral prior and is commonly used as the warm-up stage before reinforcement learning, offering a stable initialization for subsequent policy optimization.

\subsection{Reinforcement Learning}
\label{sec:grpo}

Reinforcement learning (RL) enables optimizing agent behavior via reward signals beyond token-level correctness.
A representative algorithm is Group Relative Policy Optimization (GRPO)~\cite{shao2024deepseekmath,liu2024deepseek}, which estimates advantages from group-relative comparisons without requiring a learned value function.
Given a prompt $x$, the policy $\pi_\theta$ generates a group of $G$ responses $\{y_1, \ldots, y_G\}$, each scored by a reward function $R(\cdot)$.
The advantage is estimated as:
\begin{equation}\label{eq:adv}
    A_i = \frac{R(y_i) - \text{mean}(\{R(y_j)\}_{j=1}^{G})}{\text{std}(\{R(y_j)\}_{j=1}^{G})}.
\end{equation}
The policy is updated via a clipped surrogate objective with a KL regularization term:
\begin{equation}\label{eq:grpo}
\begin{aligned}
    \mathcal{L}_{\text{GRPO}} = -\mathbb{E}\Big[ & \min\big(r_i A_i,\; \text{clip}(r_i, 1\!-\!\epsilon, 1\!+\!\epsilon) A_i\big) \\
    & - \beta \, D_{\text{KL}}(\pi_\theta \| \pi_{\text{ref}})\Big]
\end{aligned}
\end{equation}
where $r_i = \pi_\theta(y_i|x) / \pi_{\text{old}}(y_i|x)$ is the importance ratio and $\beta$ controls the deviation from the reference policy.
Recent work~\citep{Zhang2025SeeTA,wang2025customer,zhang2026shopr} has applied RL to user simulation, but their reward design still centers on step-wise imitation signals (\eg, whether each predicted action decision exactly matches the logged behavior).

However, how to equip agents with human-like memory for maintaining coherent context over multi-turn sessions, and how to optimize trajectory-level behavioral fidelity beyond single-action correctness, remain largely unaddressed.

\section{\ours{}}
\label{sec:recverse}

\subsection{Overview}
\label{sec:overview}

\begin{figure*}[t]
  \centering
  \includegraphics[width=\textwidth]{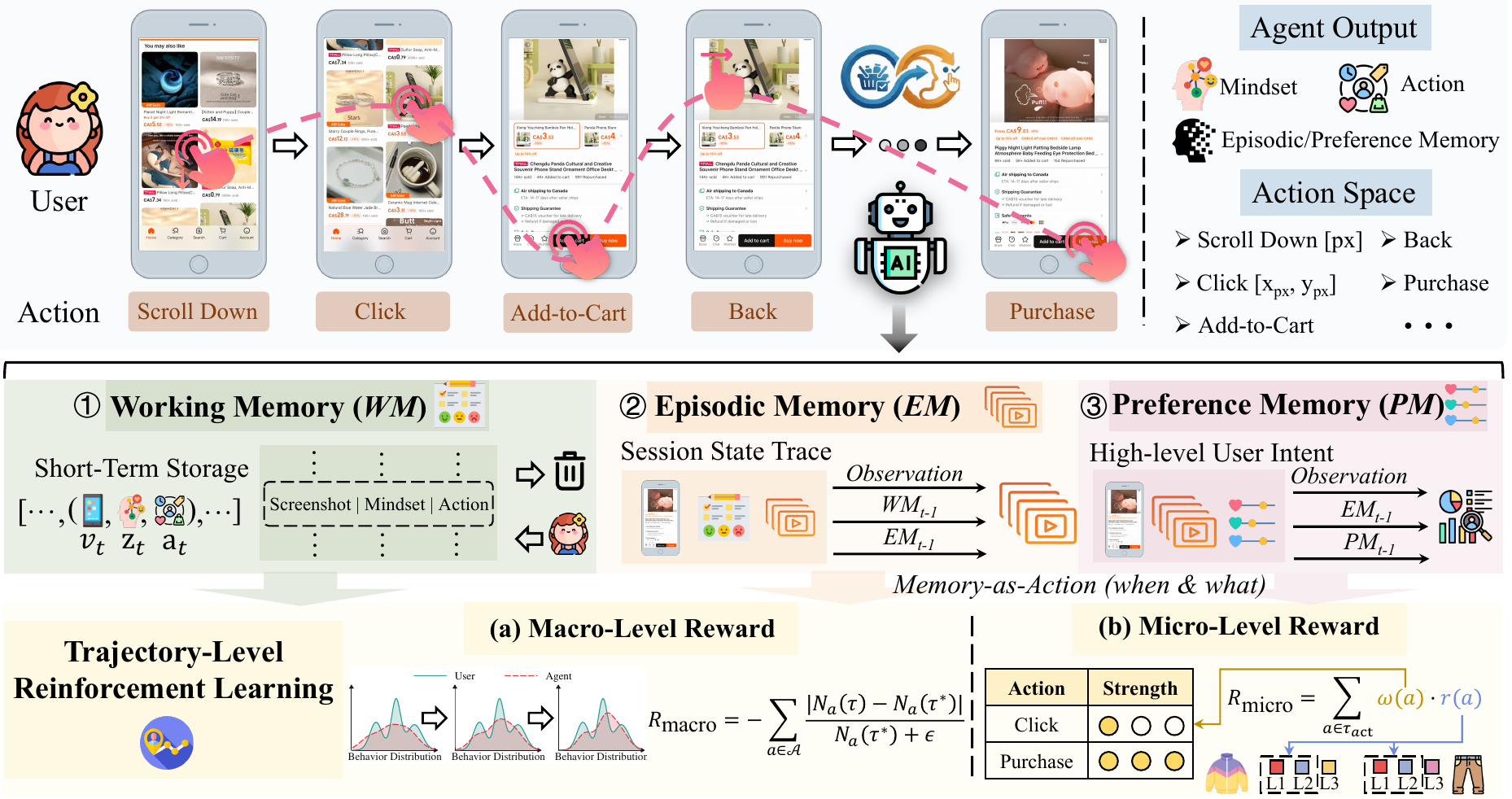}
  \vspace{-20pt}
  \caption{Overview of \ours{}. At each step, the agent perceives a page screenshot, reads from its multi-level memory, and outputs both an environment action and memory updates. The memory hierarchy comprises: \textbf{\ding{172}} \emph{Working Memory} for transient visuospatial context, \textbf{\ding{173}} \emph{Episodic Memory} for session-level behavioral traces, and \textbf{\ding{174}} \emph{Preference Memory} for distilled high-level user intent. Memory updates are themselves part of the action space (\emph{memory-as-action}), learned jointly with environment actions. The agent is optimized end-to-end via trajectory-level RL with two complementary signals: \textbf{(a)} \emph{Macro-Level Reward} for aligning aggregate item-related behavior distributions, and \textbf{(b)} \emph{Micro-Level Reward} for intent-weighted category alignment.}
  \label{fig:framework}
\end{figure*}

The key insight behind \ours{} is that faithful user simulation requires two capabilities that existing agents lack~\citep{wang2025user,Zhang2025SeeTA}: \emph{structured memory} that evolves meaningfully over a long session, and \emph{trajectory-level optimization} that looks beyond individual actions.

As illustrated in Figure~\ref{fig:framework}, at each timestep $t$, the agent perceives the current visual observation $o_t$, reads from the memory state $\mathcal{M}_{t-1}$ accumulated over previous steps, and jointly produces an internal reasoning $z_t$ (reflecting the user's mindset), an environment action $a_t$, and a memory generation $m_t$:
\begin{equation}
    z_t, a_t, m_t = \pi_\theta(o_t, \mathcal{M}_{t-1}),
\end{equation}
where $m_t$ transitions the memory state from $\mathcal{M}_{t-1}$ to $\mathcal{M}_t$.
The memory state $\mathcal{M}_t$ consists of three components operating at different cognitive granularities: a \textbf{\emph{Working Memory}} $\mathcal{W}_t$ that maintains recent visuospatial context as a short-term visual workspace, an \textbf{\emph{Episodic Memory}} $\mathcal{E}_t$ that tracks in-session behavioral context, and a \textbf{\emph{Preference Memory}} $\mathcal{P}_t$ that captures high-level user intent (\S\ref{sec:memory}).
The agent is first warm-started with imitation learning on real user trajectories, then further optimized via trajectory-level RL (\S\ref{sec:rl}).
Notably, memory updates are incorporated into the agent's action space alongside environment interactions, allowing the model to implicitly learn effective memorization strategies.

\subsection{Cognitive-Inspired Hierarchical Memory}
\label{sec:memory}

Real users do not process browsing history as a flat sequence~\citep{atkinson1968human,tulving1985many}.
They maintain different types of cognitive states at different temporal granularities: vivid short-term impressions, session-level behavioral context, and stable personal preferences.
We mirror this structure with three memory components as detailed below.

\subsubsection{\textbf{Working Memory $\mathcal{W}_t$}}
Working Memory models the visuospatial sketchpad of human cognition by maintaining a capacity-limited FIFO record of the most recent $K$ browsing steps:
\begin{equation}
    \mathcal{W}_t = \{(v_{t-K+1}, z_{t-K+1}, a_{t-K+1}), \ldots, (v_t, z_t, a_t)\},
\end{equation}
where $v_i$ denotes the visual impression retained from the viewport at step $i$, $z_i$ captures the user's current mindset (\eg, what items attract attention, what attributes are being compared, and what concerns drive the next move), and $a_i$ is the corresponding action.
When a new entry is appended, the earliest one is evicted, reflecting the capacity-limited and transient nature of human visual persistence~\citep{atkinson1968human,baddeley2000episodic,baddeley1974hitch}. This localized visual workspace grounds immediate decisions in recent perceptual context without exposing the agent to the full browsing history, which would overwhelm the context window and degrade performance~\citep{liu2024lostmiddle,wang2023longmem}.

\subsubsection{\textbf{Episodic Memory $\mathcal{E}_t$}}
Episodic Memory maintains a textual record of in-session interaction events, preserving the agent's behavioral context and ensuring coherence across pages.
Since not every step warrants a memory write, we treat memory writing as part of the action space: at each step $t$, the agent learns to emit entries selectively, deciding both \emph{when} and \emph{what} to record.
Formally, at step $t$ the agent may generate an entry $e_t$:
\begin{equation}
    \mathcal{E}_t = \mathcal{E}_{t-1} \cup \{e_t\}, \quad e_t = f_\theta(o_t, \mathcal{W}_{t-1}, \mathcal{E}_{t-1})
\end{equation}
where $f_\theta$ denotes the episodic update operator induced by the same policy model.
The decision of whether to emit $e_t$ at a given step is optimized via reinforcement learning rather than triggered by hand-crafted rules.
This serves two purposes: (i) it keeps the agent aware of its own behavioral trajectory without requiring the full visual history in context, and (ii) it provides factual grounding for Preference Memory to reason over and distill user preferences.

\subsubsection{\textbf{Preference Memory $\mathcal{P}_t$}}
Preference Memory captures the agent's higher-level understanding of user intent, distilled from observations, past actions, and accumulated episodic records:
\begin{equation}
    \mathcal{P}_t = g_\theta(o_t, \mathcal{E}_{t-1}, \mathcal{P}_{t-1}).
\end{equation}
Here, $g_\theta$ denotes the preference update operator induced by the same policy model, parallel to $f_\theta$ for episodic recording.
Like Episodic Memory, Preference Memory learns when to revise its state and what content to generate, rather than following a fixed workflow.
These updates operate at a higher level of shopping interest abstraction: rather than recording \emph{what happened}, Preference Memory distills \textbf{\emph{what the user wants}}, including preferred attributes, comparison criteria, disliked attributes, and purchase intent (\eg, ``user prefers red dresses in a low price range'').

\noindent
\paragraph{\textbf{Memory Synergy.}}
The three memory levels play complementary roles in contextual abstraction: Working Memory $\mathcal{W}_t$ retains recent perceptual context for immediate grounding, Episodic Memory $\mathcal{E}_t$ organizes interaction history into session-level events, and Preference Memory $\mathcal{P}_t$ distills these events into evolving user preferences.
At decision time, conditioning on the full hierarchy allows the agent to integrate local visual evidence, session continuity, and long-horizon preference modeling.

\subsection{Trajectory-Aligned RL}
\label{sec:rl}

Faithful user simulation is inherently trajectory-based: a policy should reproduce realistic interaction patterns and preference-driven decisions across a complete browsing session, rather than merely fit the next logged actions.
To this end, we warm-start the policy with imitation learning (\S\ref{sec:sft}) and refine it with GRPO (\S\ref{sec:grpo}) using trajectory-level feedback.
For each training session, the agent performs $G$ multi-turn rollouts in the environment, generating complete browsing trajectories including reasoning, actions, and memory updates.
We adopt a \textbf{\emph{memory-as-action}} formulation: updates to $\mathcal{E}_t$ and $\mathcal{P}_t$ are part of the agent's action space, so the agent learns \emph{when} and \emph{what} to memorize through the same reward signal that improves trajectory realism, rather than relying on hand-crafted heuristics or workflows.

\paragraph{\textbf{Reward Shaping.}}
As argued in \S\ref{sec:intro}, step-wise imitation alone is insufficient for evaluating simulation quality.
We decompose session-level alignment into two complementary reward signals:
\begin{itemize}[leftmargin=1.2em, itemsep=1pt, topsep=2pt, label=$\diamond$]
    \item \textbf{\emph{Macro-level reward}} $R_{\text{macro}}$: encourages the aggregate distribution of item-related behaviors (\eg, click, add-to-cart, purchase) to match that of real users, discouraging unrealistic patterns such as over-exploration or excessive passivity.
    \item \textbf{\emph{Micro-level reward}} $R_{\text{micro}}$: assesses how closely the agent's item-directed decisions follow the shopping intent implied by the reference trajectory.
\end{itemize}
Additionally, we include a \emph{format reward} $R_{\text{format}}$ to maintain well-structured outputs. 
In summary, the overall reward assigned to each rollout $i$ is defined as follows:
\begin{equation}
    R_i = R_{\text{macro}} + \lambda \cdot R_{\text{micro}} + R_{\text{format}}
\end{equation}
where $\lambda$ balances the two trajectory-level rewards.
The composite reward $R_i$ is then converted to a group-relative advantage $A_i$ following Eq.~\eqref{eq:adv}.
We now detail each component.

\subsubsection{\textbf{Macro-Level Reward $R_{\text{macro}}$}}
To align session-level behavioral patterns, we penalize the distributional gap between the agent's and real users' item-related actions.
Specifically, for a generated trajectory $\tau$ and the corresponding real trajectory $\tau^*$:
\begin{equation}
    R_{\text{macro}}(\tau) = -\sum_{a \in \mathcal{A}} \frac{|N_a(\tau) - N_a(\tau^*)|}{N_a(\tau^*) + \epsilon}
\end{equation}
where $\mathcal{A}$ denotes the set of item-related action types used for reward computation, excluding pure navigation such as scrolling and backtracking; $N_a(\cdot)$ counts occurrences of type $a$, and $\epsilon$ is a smoothing constant to guarantee numerical stability (we set $\epsilon = 0.1$).
This formulation emphasizes distributional agreement, suppressing unrealistic patterns such as over-exploration or excessive passivity.

\subsubsection{\textbf{Micro-Level Reward $R_{\text{micro}}$}}
While macro reward captures aggregate behavior patterns, it does not assess whether item-related decisions follow the user's shopping intent.
The micro-level reward addresses this by measuring intent-weighted category alignment over item-directed decisions.
Specifically, let $\tau_{\text{item}} \subseteq \tau$ denote the set of actions associated with concrete items:
\begin{equation}\label{eq:micro_reward}
    R_{\text{micro}}(\tau) = \sum_{a \in \tau_{\text{item}}} w(a) \cdot r(a)
\end{equation}
where $w(a)$ is an intent-strength weight derived from empirical behavioral distributions (\eg, purchases carry higher weight than clicks).
The matching score $r(a)$ measures how well the item targeted by action $a$ aligns with real user items through a hierarchical product category system (\eg, clothing $\rightarrow$ shoes $\rightarrow$ sandals):
\begin{equation}
    r(a) = \max_{y \in \mathcal{I}^*} \frac{1}{L} \sum_{l=1}^{L} \mathbb{I}(c_l[x_a] = c_l[y])
\end{equation}
where $x_a$ is the item associated with action $a$, $\mathcal{I}^*$ is the ground-truth item set from the reference trajectory, $c_l[\cdot]$ denotes the $l$-th level category label, $L$ is the depth of the category hierarchy, and $\mathbb{I}(\cdot)$ is the indicator function.
Compared with exact item matching, this multi-level category scheme better accommodates diverse plausible choices in e-commerce scenarios.
It also provides denser reward signals for agentic reinforcement learning, mitigating the sparse-gradient issue in GRPO optimization~\cite{yu2026dapo,chu2025gpg,he2026advantage}.

\subsubsection{\textbf{Format Reward $R_{\text{format}}$}}
The format reward verifies two basic validity conditions: whether the agent's output can be parsed into the required schema, and whether the proposed environment action is executable under the current interface context.
We assign a positive reward only when both conditions are satisfied, and zero otherwise, encouraging the policy to maintain reasonable output structure and execution validity during exploration phases.

\section{Experiments}
\label{sec:exp}

\begin{table}[t]
\centering
\caption{Overview statistics of USB, organized by trajectory, item, and user dimensions, respectively.}
\label{tab:dataset_stats}
\vspace{-5pt}
\renewcommand{\arraystretch}{1.0}
\begin{tabular}{llr}
\toprule
 & \textbf{Statistic} & \textbf{Value} \\
\midrule
\multirow{5}{*}{\rotatebox{90}{\scriptsize\textbf{Trajectory}}}
& \# Trajectories & 5,274 \\
& \# Action types & 8 \\
& \# Actions & 69,842 \\
& Avg. length (steps) & 13.24 \\
\midrule
\multirow{2}{*}{\rotatebox{90}{\scriptsize\textbf{Item}}}
& \# Items & 90,095 \\
& \# L1 / L2 / L3 categories & 41 / 517 / 2,256 \\
\midrule
\multirow{3}{*}{\rotatebox{90}{\scriptsize\textbf{User}}}
& \# Users & 5,222 \\
& \# Profile attributes & 5 \\
& Avg. click / purchase history & 356.48 / 10.36 \\
\bottomrule
\end{tabular}
\vspace{-10pt}
\end{table}

\begin{figure*}[t]
  \centering
  \includegraphics[width=\linewidth]{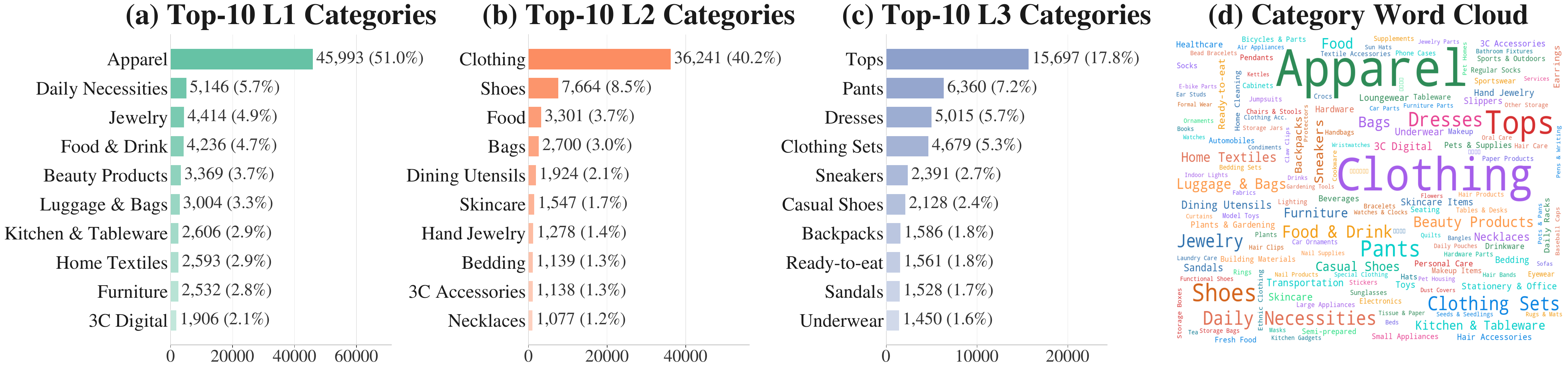}
  \vspace{-15pt}
  \caption{Category distribution analysis of USB. \textbf{(a)}, \textbf{(b)}, \textbf{(c)} show the top-10 categories at the \textbf{L1}, \textbf{L2}, and \textbf{L3} levels of the product taxonomy, respectively. \textbf{(d)} displays a word cloud of sampled categories, with font size proportional to frequency.}
  \label{fig:category_dist}
  \vspace{-5pt}
\end{figure*}

\subsection{Experimental Setup}
\label{sec:setup}

\subsubsection{\textbf{Dataset}}
We introduce \textbf{USB} (\textbf{U}ser \textbf{S}imulation \textbf{B}enchmark), a large-scale e-commerce dataset collected from a major East Asian marketplace.
USB provides complete GUI-grounded browsing trajectories with real-world user-item interactions, user profiles, and product metadata.
Beyond static trajectory logs, USB supports an interactive visual environment where agents can execute actions and receive corresponding real page observations as feedback, enabling online multi-turn RL training and evaluation.

\paragraph{\textbf{Dataset Statistics.}}
USB contains \textbf{5,274} browsing trajectories, each consisting of page-level screenshots paired with timestamped user actions.
The action space covers \textbf{8} types: \emph{scroll down}, \emph{scroll up}, \emph{click}, \emph{enter detail page}, \emph{go back}, \emph{add-to-cart}, \emph{purchase}, and \emph{terminate}.
Products are organized in a \textbf{3-tier} category taxonomy, whose distribution is visualized in Figure~\ref{fig:category_dist} (see Appendix~\ref{app:dataset} for detailed analysis).
Each user is associated with a profile containing demographic attributes (\eg, age, gender) and recent interaction history (\eg, recently clicked and purchased items), providing rich personalization information.
Key statistics are summarized in Table~\ref{tab:dataset_stats}; further details on dataset construction are in Appendix~\ref{app:dataset}.

\paragraph{\textbf{Comparison with Existing Benchmarks.}}
As shown in Table~\ref{tab:dataset_compare}, prior benchmarks each lack one or more capabilities essential for faithful user simulation.
Datasets like Amazon~\citep{hou2024bridging} and MovieLens~\citep{harper2015movielens} only contain single-type interactions (ratings or reviews) without visual context.
More recent efforts such as Qilin~\citep{chen2025qilin} and OmniBehavior~\citep{chen2026towards} introduce multiple action types but still lack real interface observations.
While OPeRA~\citep{wang2025opera} provides GUI trajectories, it is limited to static offline logs, lacking an interactive environment for dynamic exploration.
\textit{To the best of our knowledge, USB is the first user simulation benchmark that simultaneously offers GUI visual trajectories, diverse action types, user profiles, and an interactive environment for online multi-turn agentic reinforcement learning.}

\begin{table}[t]
\centering
\caption{Feature comparison of user simulation benchmarks. \underline{Visual Traj.}: provides GUI-level screenshots as observations; \underline{Diverse Actions}: supports multiple action types beyond ratings; \underline{User Profile}: includes user attributes for personalized simulation; \underline{Multi-Turn Agentic RL}: supports multi-turn agentic RL with interactive visual feedback.}
\vspace{-5pt}
\label{tab:dataset_compare}
\renewcommand{\arraystretch}{1.05}
\begin{tabular}{l@{\hspace{8pt}}c@{\hspace{6pt}}c@{\hspace{6pt}}c@{\hspace{6pt}}c}
\toprule
\textbf{Benchmark} & \makecell{\textbf{Visual}\\\textbf{Traj.}} & \makecell{\textbf{Diverse}\\\textbf{Actions}} & \makecell{\textbf{User}\\\textbf{Profile}} & \makecell{\textbf{Multi-Turn}\\\textbf{Agentic RL}} \\
\midrule
Amazon~\cite{hou2024bridging} & \textcolor{red!70!black}{\ding{55}} & \textcolor{red!70!black}{\ding{55}} & \textcolor{red!70!black}{\ding{55}} & \textcolor{red!70!black}{\ding{55}} \\
MovieLens~\cite{harper2015movielens} & \textcolor{red!70!black}{\ding{55}} & \textcolor{red!70!black}{\ding{55}} & \textcolor{green!50!black}{\ding{51}} & \textcolor{red!70!black}{\ding{55}} \\
RecBench+~\cite{huang2026towards} & \textcolor{red!70!black}{\ding{55}} & \textcolor{red!70!black}{\ding{55}} & \textcolor{green!50!black}{\ding{51}} & \textcolor{red!70!black}{\ding{55}} \\
Qilin~\cite{chen2025qilin} & \textcolor{red!70!black}{\ding{55}} & \textcolor{green!50!black}{\ding{51}} & \textcolor{green!50!black}{\ding{51}} & \textcolor{red!70!black}{\ding{55}} \\
OmniBehavior~\cite{chen2026towards} & \textcolor{red!70!black}{\ding{55}} & \textcolor{green!50!black}{\ding{51}} & \textcolor{green!50!black}{\ding{51}} & \textcolor{red!70!black}{\ding{55}} \\
OPeRA~\cite{wang2025opera} & \textcolor{green!50!black}{\ding{51}} & \textcolor{green!50!black}{\ding{51}} & \textcolor{green!50!black}{\ding{51}} & \textcolor{red!70!black}{\ding{55}} \\
\midrule
\textbf{USB (Ours)} & \textcolor{green!50!black}{\ding{51}} & \textcolor{green!50!black}{\ding{51}} & \textcolor{green!50!black}{\ding{51}} & \textcolor{green!50!black}{\ding{51}} \\
\bottomrule
\end{tabular}
\vspace{-5pt}
\end{table}

\subsubsection{\textbf{Baselines}}

We compare \ours{} with representative text-based methods (RecAgent, Agent4Rec, OPeRA, AlignUSER, Shop-R1, and Customer-R1) and GUI-based methods (A/B Agent and STA), covering training-free, IL, and RL settings.
Detailed baseline descriptions are provided in Appendix~\ref{app:baseline_details}.

\subsubsection{\textbf{Evaluation Metrics}}
We evaluate simulation quality from two complementary perspectives (\textit{behavioral fidelity} and \textit{intent consistency}), with formal definitions provided in Appendix~\ref{app:metrics}.

\paragraph{\textbf{Behavioral Fidelity Metrics.}}
These metrics measure how closely the agent's behavioral statistics match real user distributions.
The goal is \emph{minimal divergence} from ground-truth rather than maximizing or minimizing any single value.
We report \textbf{ATL} (Average Trajectory Length), \textbf{CTR} (Click-Through Rate), \textbf{IPVR} (Item Page View Rate), \textbf{ACR} (Add-to-Cart Rate), and \textbf{CVR} (Conversion Rate).

\paragraph{\textbf{Intent Consistency Metrics.}}
These metrics assess whether the agent's interaction decisions align with real user intent at the item and category levels.
All metrics can be computed jointly over all active actions or disaggregated by specific sub-action type (\eg, click only, add-to-cart).
At the item level, we report \textbf{Hit Rate (HR)}, \textbf{Precision (P)}, \textbf{Recall (R)}, and \textbf{F1} based on exact item matching.
At the category level, we define a similarity function based on the shared prefix depth in our 3-level taxonomy, and compute \textbf{Category Precision (CP)}, \textbf{Category Recall (CR)}, and their harmonic mean \textbf{Hierarchical Category Overlap (HCO)}.

\subsubsection{\textbf{Implementation Details}}
We use Qwen3.5-2B~\cite{qwen3.5} as the backbone for all trainable models and conduct distributed training with Megatron-LM framework~\cite{shoeybi2019megatron}. Detailed training hyperparameters and implementation settings are provided in Appendix~\ref{app:implementation_details}.

\subsection{Overall Performance}
\label{sec:overall_performance}

\begin{figure}[t]
  \centering
  \includegraphics[width=\columnwidth]{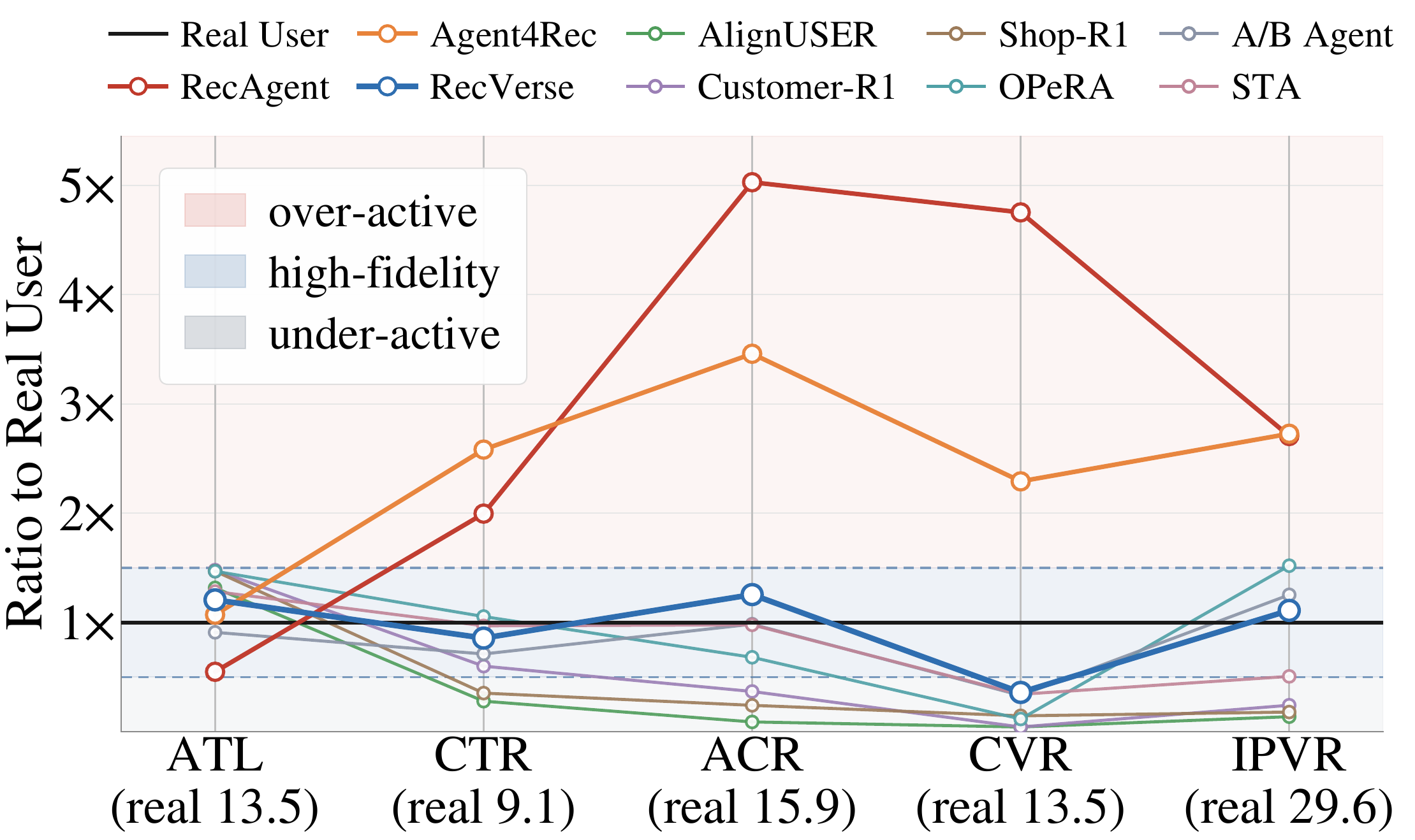}
  \vspace{-15pt}
  \caption{Behavioral fidelity comparison normalized by real-user statistics. Each point reports the ratio between a method's behavioral statistic and the corresponding real-user value. The shaded regions distinguish over-active, high-fidelity, and under-active behavioral regimes.}
  \label{fig:behavior_fidelity}
  \vspace{-15pt}
\end{figure}

Table~\ref{tab:main_results} summarizes the overall results, while Figure~\ref{fig:behavior_fidelity} visualizes behavioral fidelity by normalizing each statistic against its real-user reference. We examine intent consistency jointly with behavioral fidelity, as high item/category overlap can be trivially inflated by an over-active interaction policy rather than reflecting faithful user simulation. RecAgent and Agent4Rec exemplify this pitfall: although their intent-based scores appear competitive, their ACR, CVR, and IPVR are substantially higher than those of real users, and both consistently reside in the over-active regime across multiple behavioral dimensions (as shown in Figure~\ref{fig:behavior_fidelity}). Their hit-based performance gains therefore arise from over-interaction rather than realistic browsing dynamics. Methods adapted to logged behavior, in contrast, produce far less extreme interaction patterns, indicating that task-specific optimization helps curb degenerate exploration and promotes more faithful browsing behavior.

Among behaviorally comparable methods, \ours{} benefits from both GUI-grounded perception and trajectory-level RL. Moving from text to GUI substantially improves category-level alignment, with HCO increasing from 19.03 to 30.00 under IL and from 24.38 to 32.64 under RL, indicating that pixel-level page observations provide useful evidence beyond textual item descriptions. RL further strengthens the GUI-grounded simulator: compared with STA, the strongest GUI baseline, \ours{}-GUI with RL improves item-level F1 from 4.27 to 7.19 and HR from 5.92 to 10.45, while HCO increases from 23.11 to 32.64. Together with its proximity to the real-user high-fidelity band in Figure~\ref{fig:behavior_fidelity}, these results show that \ours{}-GUI with RL jointly achieves realistic browsing behavior and strong intent alignment across both item and category levels.

\begin{table*}[t]
\centering
\caption{Overall evaluation performance. \textbf{TF}, \textbf{IL}, and \textbf{RL} denote training-free prompting, imitation learning, and reinforcement learning, respectively. All metrics except ATL are reported as percentages. \colorbox{red!4}{Red rows} indicate methods whose behavioral statistics substantially deviate from real users due to over-active patterns, and \colorbox{blue!5}{blue rows} denote \ours{} variants. For intent consistency metrics, bold indicates the best score among behaviorally comparable methods within each modality group.}
\label{tab:main_results}
\vspace{-5pt}
\begin{tabular}{cll ccccc ccccccc c}
\toprule
& \multirow{2}{*}{\textbf{Method}} & \multirow{2}{*}{\textbf{Train}} & \multicolumn{5}{c}{\textbf{Behavioral Fidelity}} & \multicolumn{7}{c}{\textbf{Intent Consistency}} & \multirow{2}{*}{\textbf{Format}} \\
\cmidrule(lr){4-8}\cmidrule(lr){9-15}
& & & ATL & CTR & ACR & CVR & IPVR & P & R & F1 & HR & CP & CR & HCO & \\
\midrule
Real & Real User & -- & 13.47 & 9.08 & 15.93 & 13.54 & 29.60 & -- & -- & -- & -- & -- & -- & -- & -- \\
\midrule
\multirow{8}{*}{\rotatebox[origin=c]{90}{\textbf{Text}}}
& \cellcolor{red!4}\emph{RecAgent}$^\dagger$ & \cellcolor{red!4}TF & \cellcolor{red!4}7.39 & \cellcolor{red!4}18.12 & \cellcolor{red!4}80.09 & \cellcolor{red!4}64.34 & \cellcolor{red!4}80.09 & \cellcolor{red!4}7.69 & \cellcolor{red!4}5.88 & \cellcolor{red!4}6.67 & \cellcolor{red!4}8.09 & \cellcolor{red!4}41.06 & \cellcolor{red!4}35.84 & \cellcolor{red!4}38.27 & \cellcolor{red!4}100.00 \\
& \cellcolor{red!4}\emph{Agent4Rec}$^\dagger$ & \cellcolor{red!4}TF & \cellcolor{red!4}14.42 & \cellcolor{red!4}23.44 & \cellcolor{red!4}55.09 & \cellcolor{red!4}31.02 & \cellcolor{red!4}80.70 & \cellcolor{red!4}9.36 & \cellcolor{red!4}12.54 & \cellcolor{red!4}10.72 & \cellcolor{red!4}16.96 & \cellcolor{red!4}43.90 & \cellcolor{red!4}50.13 & \cellcolor{red!4}46.81 & \cellcolor{red!4}93.74 \\
& \emph{OPeRA} & TF & 19.78 & 9.57 & 10.85 & 1.58 & 44.97 & 3.16 & 2.14 & 2.55 & 3.16 & 24.52 & 20.90 & 22.57 & 99.88 \\
& \emph{AlignUSER} & IL & 17.73 & 2.54 & 1.43 & 0.59 & 4.08 & 2.94 & 2.27 & 2.56 & 3.75 & 20.06 & 18.73 & 19.37 & 100.00 \\
& \emph{Shop-R1} & RL & 19.79 & 3.21 & 3.85 & 1.97 & 5.33 & 1.78 & 1.35 & 1.53 & 1.78 & 8.55 & 7.29 & 7.87 & 100.00 \\
& \emph{Customer-R1} & RL & 19.92 & 5.45 & 5.87 & 0.59 & 7.18 & 1.87 & 1.31 & 1.55 & 1.97 & 12.33 & 11.20 & 11.74 & 100.00 \\
\cdashline{2-16}
& \cellcolor{blue!5}\ours{}-Text & \cellcolor{blue!5}IL & \cellcolor{blue!5}13.81 & \cellcolor{blue!5}3.25 & \cellcolor{blue!5}9.80 & \cellcolor{blue!5}4.57 & \cellcolor{blue!5}6.31 & \cellcolor{blue!5}\textbf{3.75} & \cellcolor{blue!5}\textbf{3.14} & \cellcolor{blue!5}\textbf{3.42} & \cellcolor{blue!5}\textbf{4.34} & \cellcolor{blue!5}20.02 & \cellcolor{blue!5}18.13 & \cellcolor{blue!5}19.03 & \cellcolor{blue!5}100.00 \\
& \cellcolor{blue!5}\ours{}-Text & \cellcolor{blue!5}RL & \cellcolor{blue!5}15.61 & \cellcolor{blue!5}4.70 & \cellcolor{blue!5}10.98 & \cellcolor{blue!5}7.59 & \cellcolor{blue!5}7.64 & \cellcolor{blue!5}3.52 & \cellcolor{blue!5}2.96 & \cellcolor{blue!5}3.21 & \cellcolor{blue!5}3.94 & \cellcolor{blue!5}\textbf{25.85} & \cellcolor{blue!5}\textbf{23.06} & \cellcolor{blue!5}\textbf{24.38} & \cellcolor{blue!5}98.99 \\
\midrule
\multirow{4}{*}{\rotatebox[origin=c]{90}{\textbf{GUI}}}
& \emph{A/B Agent} & TF & 12.25 & 6.46 & 15.68 & 4.50 & 37.08 & 4.21 & 4.21 & 4.21 & 5.52 & 22.34 & 20.99 & 21.64 & 98.31 \\
& \emph{STA} & RL & 17.28 & 8.80 & 15.58 & 4.64 & 14.99 & 4.73 & 3.90 & 4.27 & 5.92 & 24.46 & 21.89 & 23.11 & 99.35 \\
\cdashline{2-16}
& \cellcolor{blue!5}\ours{}-GUI & \cellcolor{blue!5}IL & \cellcolor{blue!5}15.29 & \cellcolor{blue!5}5.70 & \cellcolor{blue!5}14.07 & \cellcolor{blue!5}5.46 & \cellcolor{blue!5}24.62 & \cellcolor{blue!5}5.49 & \cellcolor{blue!5}5.07 & \cellcolor{blue!5}5.27 & \cellcolor{blue!5}7.10 & \cellcolor{blue!5}31.51 & \cellcolor{blue!5}28.64 & \cellcolor{blue!5}30.00 & \cellcolor{blue!5}99.99 \\
& \cellcolor{blue!5}\ours{}-GUI & \cellcolor{blue!5}RL & \cellcolor{blue!5}16.25 & \cellcolor{blue!5}7.78 & \cellcolor{blue!5}19.99 & \cellcolor{blue!5}4.87 & \cellcolor{blue!5}32.86 & \cellcolor{blue!5}\textbf{7.41} & \cellcolor{blue!5}\textbf{6.98} & \cellcolor{blue!5}\textbf{7.19} & \cellcolor{blue!5}\textbf{10.45} & \cellcolor{blue!5}\textbf{34.51} & \cellcolor{blue!5}\textbf{30.95} & \cellcolor{blue!5}\textbf{32.64} & \cellcolor{blue!5}99.84 \\
\bottomrule
\end{tabular}
\end{table*}

\subsection{Ablation Study}
\label{sec:ablation_study}

\begin{figure}[t]
  \centering
  \includegraphics[width=\columnwidth]{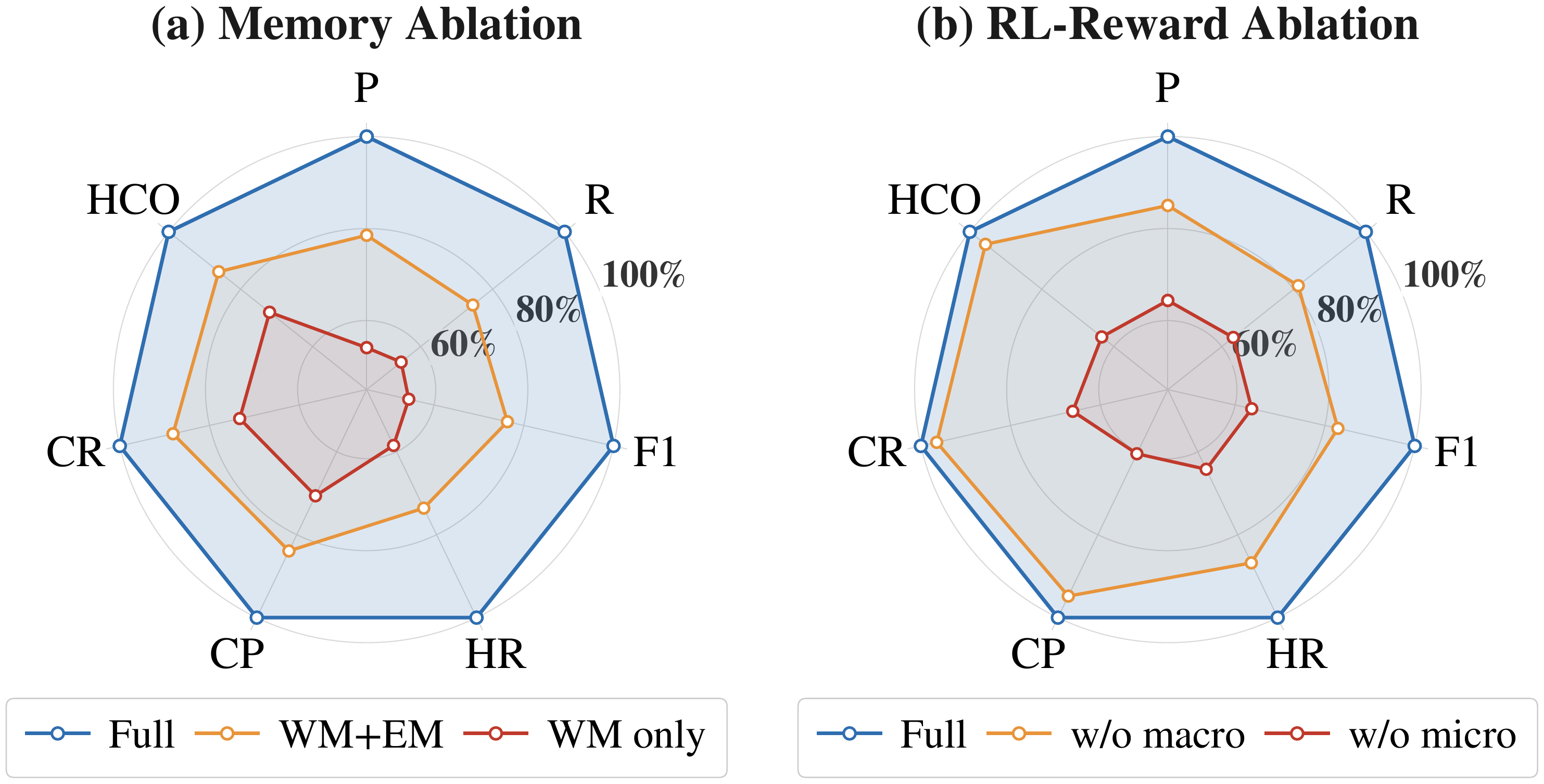}
  \vspace{-10pt}
  \caption{Ablation analysis of \ours{}. \textbf{(a)} Memory ablation compares the full memory hierarchy with variants that remove Preference Memory (PM) or retain only Working Memory (WM). \textbf{(b)} RL-reward ablation evaluates the contributions of macro-level and micro-level rewards. Scores are normalized relative to the full model for each metric.}
  \label{fig:ablation}
  \vspace{-10pt}
\end{figure}

To disentangle the contribution of each core design, we ablate the memory hierarchy and the trajectory-level reward components, as illustrated in Figure~\ref{fig:ablation}. On the memory side, the full model consistently surpasses variants that either drop Preference Memory (PM) or retain only Working Memory (WM), confirming that short-term perceptual context alone cannot sustain coherent shopping intent. Episodic Memory (EM) and PM play complementary roles at different levels of abstraction: the former organizes session-level events, while the latter accumulates higher-level user interests that guide subsequent decisions in the shopping scenario.

As shown in Figure~\ref{fig:ablation}\textbf{(b)}, the reward ablation further confirms the necessity of intent-aware optimization. Removing the micro-level reward leads to the largest degradation across item-level and category-level metrics, indicating that distributional behavior matching alone cannot recover fine-grained shopping intent. Removing the macro-level reward is less harmful to intent scores, but it weakens the trajectory-level constraint that prevents unrealistic interaction patterns. These results support the joint design of hierarchical memory system and trajectory-aligned rewards.

\subsection{Human Evaluation}
\label{sec:human_evaluation}

\begin{figure}[t]
  \centering
  \includegraphics[width=0.95\columnwidth]{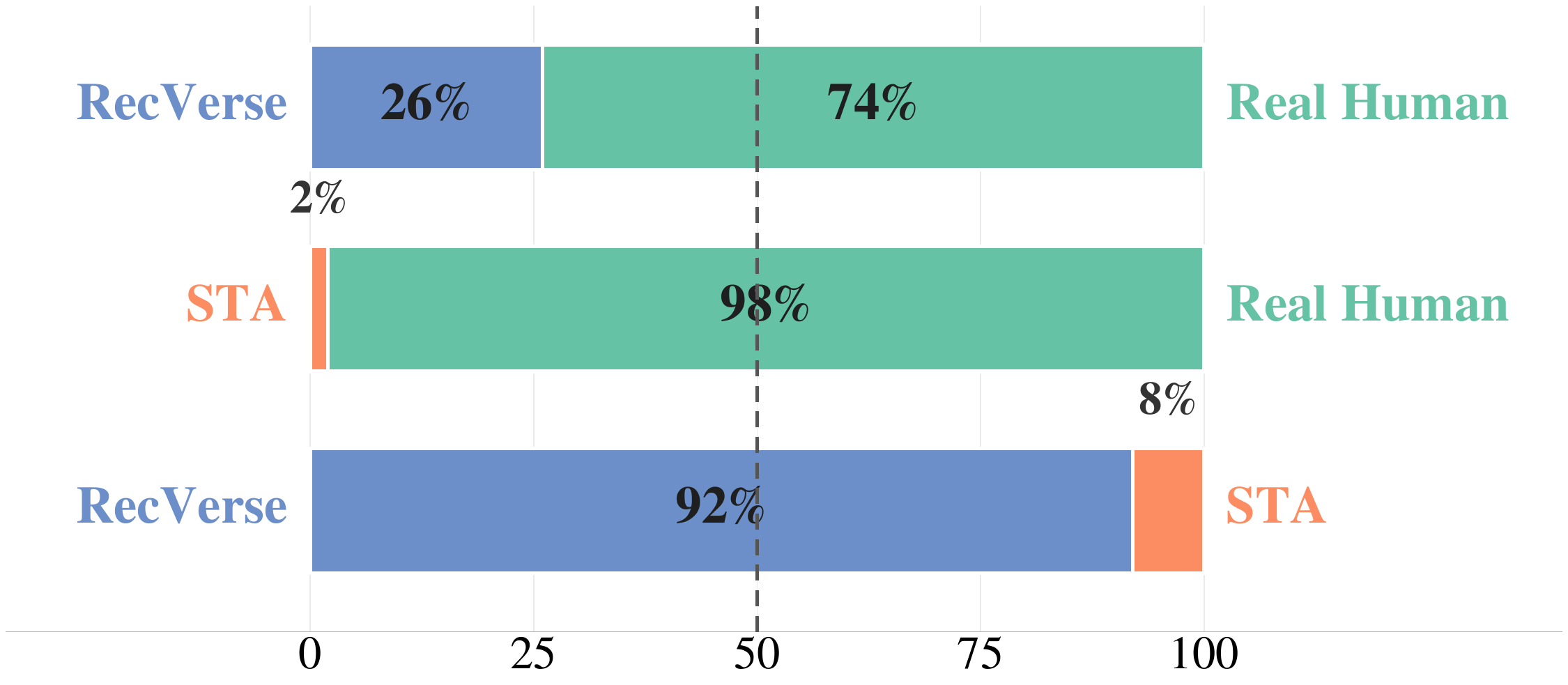}
  \vspace{-5pt}
  \caption{Human evaluation via pairwise preference judgments. Annotators compare trajectories generated by \ours{}, STA, and real users, where larger segments indicate higher preference rates for the method shown on each side.}
  \label{fig:human_eval}
  \vspace{-10pt}
\end{figure}

Beyond automatic metrics, which capture objective distributional alignment, we further conduct a pairwise human preference evaluation to assess the perceived realism of agent-simulated trajectories relative to real-user ones; the resulting labels show strong inter-annotator agreement (Fleiss' $\kappa=0.834$). As shown in Figure~\ref{fig:human_eval}, annotators can still distinguish real user trajectories from \ours{}, preferring real users in 74\% of comparisons. However, this gap is substantially smaller than that of STA, for which real users are preferred in 98\% of comparisons. When directly comparing the two simulators, \ours{} is preferred over STA in 92\% of cases, indicating that our proposed memory and trajectory-level optimization produce behavior that is more aligned with human judgment. However, the remaining gap to real users also suggests that fully human-like simulation still requires deeper personalization and richer modeling of individual browsing preferences in future work.

\subsection{Further Analysis}
\label{sec:further_analysis}

Beyond the main comparison and ablation study, we further examine two factors that affect \ours{}: the micro-level reward coefficient and model scale. We focus on these quantitative analyses in the main text, and provide a qualitative case study in Appendix~\ref{app:case_study}.

\subsubsection{\textbf{Reward Sensitivity}}
Table~\ref{tab:lambda_analysis} analyzes the micro-level reward coefficient $\lambda$. When $\lambda$ is small, the simulator stays closer to the real-user trajectory length: $\lambda=1$ gives the lowest $\Delta$ATL of 0.26. However, this setting provides weak intent guidance, with F1, HR, and HCO remaining at 3.65, 5.13, and 27.97. Increasing $\lambda$ steadily strengthens intent alignment. At $\lambda=1000$, F1 rises to 7.19, HR reaches 10.45, and HCO improves to 32.64, although the trajectory-length deviation also grows to 2.78. This suggests that $\lambda$ controls how strongly the policy prioritizes fine-grained intent alignment, while $\Delta$ATL serves as a useful check on whether this optimization distorts trajectory-level behavior.

\begin{table}[t]
\centering
\caption{Sensitivity analysis of the reward coefficient $\lambda$. $\Delta$ATL is computed against the real-user average trajectory length.}
\label{tab:lambda_analysis}
\vspace{-5pt}
\renewcommand{\arraystretch}{1}
\begin{tabular}{ccccc}
\toprule
$\lambda$ & $\Delta$ATL$\downarrow$ & F1$\uparrow$ & HR$\uparrow$ & HCO$\uparrow$ \\
\midrule
1    & \textbf{0.26} & 3.65 & 5.13 & 27.97 \\
10   & 2.42 & 4.80 & 6.71 & 28.65 \\
100  & 1.63 & 5.31 & 6.90 & 30.89 \\
1000 & 2.78 & \textbf{7.19} & \textbf{10.45} & \textbf{32.64} \\
\bottomrule
\end{tabular}
\vspace{-10pt}
\end{table}

\subsubsection{\textbf{Scaling Analysis}}
\begin{figure}[t]
  \centering
  \includegraphics[width=\columnwidth]{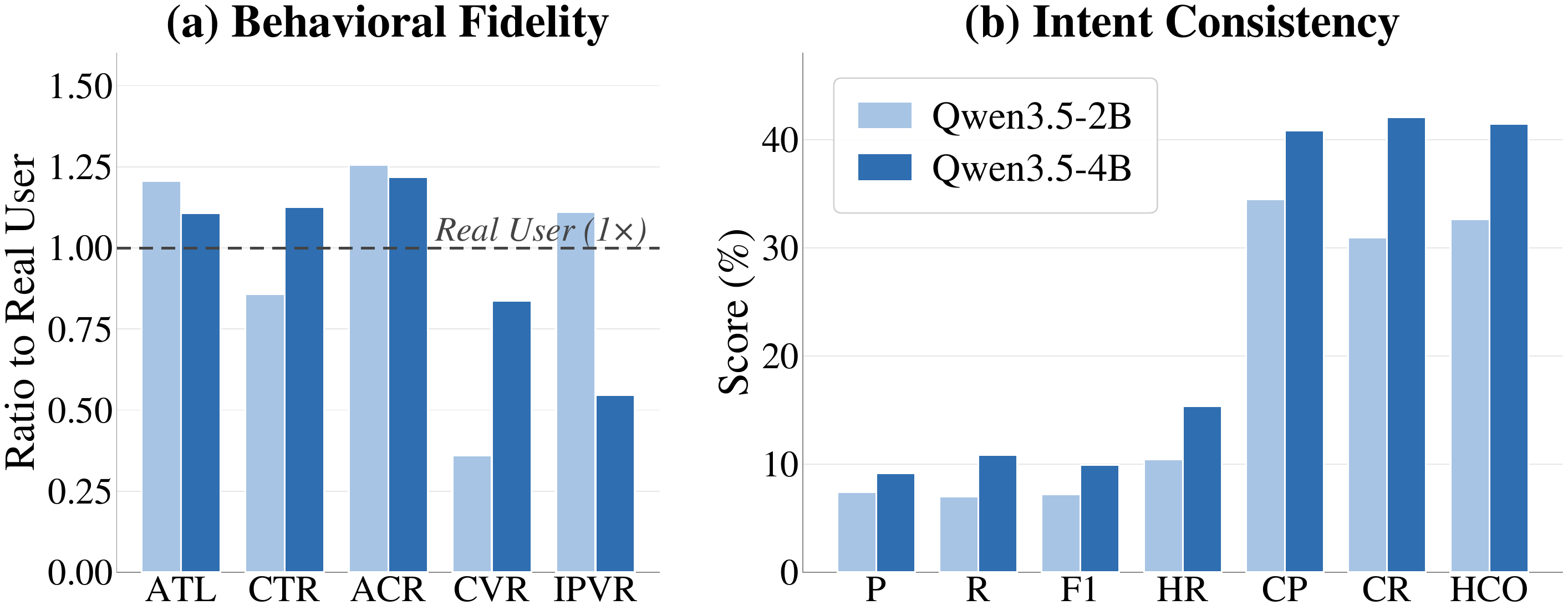}
  \vspace{-15pt}
  \caption{Scaling analysis from Qwen3.5-2B to Qwen3.5-4B on behavioral fidelity and intent consistency metrics.}
  \label{fig:scaling}
   \vspace{-10pt}
\end{figure}

Figure~\ref{fig:scaling} examines whether the proposed framework benefits from a larger backbone. On behavioral fidelity (Figure~\ref{fig:scaling}\textbf{(a)}), scaling does not simply make the agent more active. The 4B model keeps ATL close to the real-user reference and brings CTR from a slightly under-active level to around the 1$\times$ line. ACR remains mildly above the real-user rate for both models, while CVR improves substantially from a severely under-active regime in the 2B model to a much closer value in the 4B model. At the same time, IPVR becomes lower than the real-user reference, indicating that scaling improves several behavioral dimensions but does not uniformly match all browsing statistics. On intent consistency (Figure~\ref{fig:scaling}\textbf{(b)}), the improvement is more consistent: P, R, F1, and HR all increase, with the largest relative gain appearing in HR, suggesting that the larger model more reliably reaches user-relevant items. Category-level metrics also rise from the low-30\% range to above 40\%, showing stronger recovery of broader shopping interests. Overall, scaling mainly strengthens intent modeling while maintaining broadly realistic session-level behavior, though some behavioral dimensions still leave room for further calibration.

\section{Related Work}
\label{sec:related}

\ours{} draws on two lines of research: GUI agents and user behavior simulation. We briefly review each below.

\subsection{GUI Agents}
Vision-language models have driven rapid progress in agents that perceive and act on graphical interfaces directly from screenshots, spanning web~\citep{gur2024real,deng2023mind2web,zhou2024webarena,koh2024visualwebarena}, mobile~\citep{wang2024mobileagent,zhang2025appagent}, desktop~\citep{xie2024osworld}, and cross-application~\citep{trivedi2024appworld} platforms.
Foundation models such as CogAgent~\citep{hong2024cogagent}, SeeClick~\citep{cheng2024seeclick}, and UI-TARS~\citep{qin2025ui} learn pixel-level interface grounding at scale, while RL pipelines like DigiRL~\citep{bai2024digirl} further improve task success in dynamic environments through outcome-based supervision.
Despite this breadth, a common thread is the \textcolor{red!70!black}{\emph{goal-oriented}} formulation: success is measured by whether a prescribed task (\eg, ``book a flight'', ``find item X'') is completed, and supervision reduces to sparse end-task indicators~\citep{zheng2024gpt4vision}, making the agent indifferent to \emph{how} the goal is reached.
\ours{} repurposes GUI perception for \textcolor{green!70!black}{\emph{process-oriented}} user behavior simulation, where the objective is not to reach a goal but to faithfully reproduce how real users browse, hesitate, compare, and decide.
This shift fundamentally changes both the modeling requirements (long-horizon behavioral coherence, intent exploration, faithful action distributions) and the evaluation protocol (distributional fidelity, intent consistency) compared with standard GUI agent benchmarks, and motivates the memory and trajectory-level RL designs in \ours{}.

\subsection{User Behavior Simulation}
Simulating realistic user behavior underpins offline evaluation and policy training for recommender systems~\citep{ie2019recsim,mladenov2021recsimng}.
Classical simulators build probabilistic models over vectorized item features~\citep{shi2019virtual}, while LLM-based agents add natural-language reasoning and persona conditioning over textual histories~\citep{wang2025user,zhang2024generative,zhang2024agentcf,wang2025opera}. Both lines, however, ignore the visual layout that drives real browsing decisions.
A more recent line introduces explicit supervision over real trajectories, via imitation learning~\citep{bougie2026alignuser} or step-level reinforcement learning~\citep{zhang2026shopr,wang2025customer,lu2025can}.
On the visual side, STA~\citep{Zhang2025SeeTA} conditions on the current screenshot together with full action history and heuristically pruned historical HTML observations, and trains with step-level rewards, while A/B Agent~\citep{Zhang2026ExploringRS} relies on training-free prompting with hand-crafted decision modules.
Across both text-based and image-based user simulators, existing work often naively concatenates long-range history in an unstructured manner, relies on per-step action matching rather than trajectory-level objective, and, when memory is incorporated, manages with generic heuristics~\citep{shinn2023reflexion,zhong2024memorybank,wang2024voyager} instead of cognitive-inspired memory mechanisms for human browsing~\citep{atkinson1968human,tulving1985many,baddeley2000episodic}.
\ours{} addresses these gaps with a cognitive-inspired hierarchical memory whose updates are learned as part of the agent's action space, jointly optimized with a trajectory-level RL objective~\citep{shao2024deepseekmath,liu2024deepseek} that aligns simulated behavior with real users at both the distributional and intent levels.

\section{Conclusion}

We studied GUI-grounded user behavior simulation for e-commerce recommendation, targeting both realistic browsing dynamics and user intent. We introduced \ours{}, a GUI-grounded agent that combines cognitive-inspired hierarchical memory with trajectory-aligned RL to address long-horizon context modeling and step-wise imitation limitations. We also constructed USB, an interactive benchmark with real GUI trajectories, diverse actions, user profiles, and pixel-level observations. Extensive experiments show that \ours{} improves intent consistency while preserving realistic behavior, producing trajectories closer to real users than strong GUI-based baselines. We hope this work encourages more reliable and faithful user simulators for offline evaluation, counterfactual analysis, and RL-based recommender optimization. 

\bibliographystyle{ACM-Reference-Format}
\bibliography{sample-sigconf}

\clearpage

\appendix
\noindent\textbf{Appendix Overview}
\vspace{2pt}
\begin{center}
\setlength{\tabcolsep}{6pt}
\renewcommand{\arraystretch}{1.15}
\arrayrulecolor{black!25}
\begin{tabular}{|p{0.30\linewidth}|p{0.55\linewidth}|}
\hline
\rowcolor{blue!12}\textbf{Appendix} & \textbf{Title} \\
\hline
Appendix~\ref{app:metrics} & Evaluation Metric Definitions \\
\rowcolor{gray!6}\hspace{0.5em}$\hookrightarrow$ Appendix~\ref{app:behavioral_metrics} & Behavioral Fidelity Metrics \\
\hspace{0.5em}$\hookrightarrow$ Appendix~\ref{app:intent_metrics} & Intent Consistency Metrics \\
\rowcolor{gray!6}Appendix~\ref{app:dataset} & Dataset Details \\
\hspace{0.5em}$\hookrightarrow$ Appendix~\ref{app:category_dist} & Category Distribution Analysis \\
\rowcolor{gray!6}\hspace{0.5em}$\hookrightarrow$ Appendix~\ref{app:interactive_env} & Environment Construction \\
Appendix~\ref{app:exp_details} & Experimental Details \\
\rowcolor{gray!6}\hspace{0.5em}$\hookrightarrow$ Appendix~\ref{app:baseline_details} & Baseline Details \\
\hspace{0.5em}$\hookrightarrow$ Appendix~\ref{app:implementation_details} & Implementation Details \\
\rowcolor{gray!6}\hspace{0.5em}$\hookrightarrow$ Appendix~\ref{app:prompt_template} & Prompt Template \\
Appendix~\ref{app:additional_analysis} & Additional Analysis \\
\rowcolor{gray!6}\hspace{0.5em}$\hookrightarrow$ Appendix~\ref{app:micro_reward_design} & Effect of Matching Granularity \\
\hspace{0.5em}$\hookrightarrow$ Appendix~\ref{app:case_study} & Qualitative Case Study \\
\rowcolor{gray!6}Appendix~\ref{app:human_eval_interface} & Human Evaluation Interface \\
\hline
\end{tabular}
\arrayrulecolor{black}
\end{center}

\section{Evaluation Metric Definitions}
\label{app:metrics}

Since no established protocol exists for long-term user behavior simulation, we organize evaluation around two complementary questions: \ding{172} \emph{whether the simulator behaves like a real user in aggregate}, and \ding{173} \emph{whether it engages with the content a real user would}.

For the former, we calibrate against the funnel statistics that industrial recommender systems monitor in production (\eg, click-through, add-to-cart, and conversion rates), and compare them directly with real-user references instead of normalized divergences such as KL or JS, which are insensitive to absolute interaction volume and would let an over-active agent with realistic action proportions appear well aligned.
As these statistics concern action types rather than the items acted upon, we further assess intent consistency via exact item overlap and hierarchical category overlap that grants partial credit to semantically related choices.

\subsection{Behavioral Fidelity Metrics}
\label{app:behavioral_metrics}
For behavioral fidelity metrics, we compute the following rates from aggregated counts over all evaluated trajectories.
Let $N_{\text{exp}}$ denote the total number of exposed items, and let $N_a$ denote the total number of actions of type $a$. We define:
\[
    \text{CTR} = \frac{N_{\text{click}}}{N_{\text{exp}}}, \qquad
    \text{IPVR} = \frac{N_{\text{detail}}}{N_{\text{click}}}
\]
\[
    \text{ACR} = \frac{N_{\text{cart}}}{N_{\text{click}}}, \qquad
    \text{CVR} = \frac{N_{\text{purchase}}}{N_{\text{click}}}
\]

\noindent where \textbf{CTR} is Click-Through Rate, \textbf{IPVR} is Item Page View Rate after click, \textbf{ACR} is Add-to-Cart Rate after click, and \textbf{CVR} is Conversion Rate after click.
Additionally, we report \textbf{ATL} (Average Trajectory Length), \ie, the average number of steps in a trajectory.

For behavioral fidelity evaluation, the goal is \emph{NOT} to maximize or minimize any single metric, but to minimize the divergence between the agent's statistics and those of real users.
A faithful simulator should produce behavioral distributions that closely match the ground-truth: the closer the agent's metrics are to the real user values, the more realistic the simulation.

\subsection{Intent Consistency Metrics}
\label{app:intent_metrics}
Let $I_a$ and $I_u$ denote the item sets interacted with by the agent and the real user respectively.

\paragraph{\textbf{Item-Level.}}
We measure exact item overlap between the agent's and real user's interaction sets via \textbf{Hit Rate (HR)}, \textbf{Precision (P)}, \textbf{Recall (R)}, and \textbf{F1} score as follows:
\begin{equation}
    \text{HR} = \mathbb{I}(|I_a \cap I_u| > 0)
\end{equation}
\begin{equation}
    \text{P} = \frac{|I_a \cap I_u|}{|I_a|}, \qquad \text{R} = \frac{|I_a \cap I_u|}{|I_u|}, \qquad \text{F1} = \frac{2 \cdot \text{P} \cdot \text{R}}{\text{P} + \text{R}}
\end{equation}
All item-level metrics are computed per trajectory and then averaged over the test set.

\paragraph{\textbf{Category-Level.}}
We measure soft category alignment via Category Precision (CP), Category Recall (CR), and their Hierarchical Category Overlap (HCO).
We first define item similarity based on the shared prefix depth in our category hierarchy:
\begin{equation}
    \text{sim}(a, b) = \frac{|\text{LCP}(a, b)|}{L}
\end{equation}
where $\text{LCP}(a, b)$ denotes the longest common prefix in the category tree and $L$ is the maximum depth in the category tree ($L = 3$ in our experiments).
Then, we define the following metrics:
\begin{itemize}[leftmargin=1em, itemsep=2pt, label=$\diamond$]
    \item \textbf{Category Precision (CP)}:
    \begin{equation}
        \text{CP} = \frac{1}{|I_a|} \sum_{a_i \in I_a} \max_{u_j \in I_u} \text{sim}(a_i, u_j)
    \end{equation}
    \item \textbf{Category Recall (CR)}:
    \begin{equation}
        \text{CR} = \frac{1}{|I_u|} \sum_{u_j \in I_u} \max_{a_i \in I_a} \text{sim}(u_j, a_i)
    \end{equation}
    \item \textbf{Hierarchical Category Overlap (HCO)}:
    \begin{equation}
        \text{HCO} = \frac{2 \cdot \text{CP} \cdot \text{CR}}{\text{CP} + \text{CR}}
    \end{equation}
\end{itemize}

All intent metrics can be computed over all active actions or restricted to a specific sub-action type (\eg, click, add-to-cart).

\begin{table*}[t]
\centering
\caption{Condensed prompt template for rollout generation. Placeholders are filled with the user profile, historical records, current memory state, GUI observation, and executable action set at each step.}
\label{tab:prompt_template}
\small
\renewcommand{\arraystretch}{1.2}
\resizebox{\textwidth}{!}{
\begin{tabular}{p{0.11\textwidth}p{0.95\textwidth}}
\toprule
\textbf{Block} & \textbf{Template content} \\
\midrule
System role & Role-play as a real mobile e-commerce user browsing a personalized recommendation feed; predict the next natural action and the first-person mindset behind it. \\
User context & User profile \texttt{\{user\_profile\}}, historical clicks \texttt{\{click\_history\}}, historical purchases \texttt{\{buy\_history\}}, and current step \texttt{\{turn\_idx\}}. \\
Session memory & Working Memory stores recent page-level visual impressions, previous actions, and mindset within a sliding window; Episodic Memory \texttt{\{episodic\_memory\}} records current-session events and browsing state; Preference Memory \texttt{\{preference\_memory\}} stores stable reusable preferences. \\
Observation & Current page \texttt{\{page\_name\}}, current screenshot, and runtime executable action set \texttt{\{available\_actions\}}. \\
Action space & Global end action; feed-page scrolling and item clicking; product-detail actions including detail viewing, add-to-cart, purchase, and return-to-home. The prompt only exposes actions executable under the current page state. \\
Decision principle & Choose one executable action grounded in the visible content and consistent with the user profile, historical interactions, and session memory. Avoid overly deterministic behavior, including frequent clicking, purchasing, or preference updates when evidence is weak. \\
Memory update & Update Episodic Memory only for noticeable session-state changes. Update Preference Memory only when the behavior provides clear reusable preference evidence, such as purchase, add-to-cart, or in-depth detail-page browsing. \\
Output schema & Strict JSON with four fields: \texttt{mindset}, \texttt{episodic\_memory}, \texttt{preference\_memory}, and \texttt{action}. \\
\bottomrule
\end{tabular}
}
\end{table*}

\section{Dataset Details}
\label{app:dataset}

\subsection{Category Distribution Analysis}
\label{app:category_dist}
To characterize the diversity and coverage of USB, we analyze the distribution of product categories across the three levels of our taxonomy.
Figure~\ref{fig:category_dist} presents the top-10 categories at each level along with a word cloud visualization of sampled categories.

As shown in Figure~\ref{fig:category_dist}\textbf{(a)}, the dataset is concentrated in \emph{Apparel} (51.0\%), while still covering a broad range of long-tail domains such as \emph{Daily Necessities} (5.7\%), \emph{Jewelry} (4.9\%), \emph{Food \& Drink} (4.7\%), \emph{Beauty Products} (3.7\%), and \emph{Luggage \& Bags} (3.3\%).
At L2 (Figure~\ref{fig:category_dist}\textbf{(b)}), \emph{Clothing} (40.2\%) and \emph{Shoes} (8.5\%) remain the dominant apparel-related subcategories, followed by non-apparel categories such as \emph{Food} (3.7\%), \emph{Bags} (3.0\%), \emph{Dining Utensils} (2.1\%), and \emph{Skincare} (1.7\%).
The L3 distribution (Figure~\ref{fig:category_dist}\textbf{(c)}) further reveals fine-grained shopping intents, with \emph{Tops} (17.8\%) as the largest category, followed by \emph{Pants} (7.2\%), \emph{Dresses} (5.7\%), \emph{Clothing Sets} (5.3\%), as well as footwear and lifestyle categories such as \emph{Sneakers}, \emph{Casual Shoes}, \emph{Backpacks}, \emph{Ready-to-eat}, and \emph{Sandals}.
The word cloud in Figure~\ref{fig:category_dist}\textbf{(d)} provides an intuitive overview of category prevalence across all three levels, where font size corresponds to frequency.
Overall, the category distribution exhibits a natural long-tail pattern typical of real-world e-commerce platforms, ensuring that USB captures both dominant consumer interests and diverse niche shopping intents.

\subsection{Environment Construction}
\label{app:interactive_env}

USB is built upon the homepage recommendation feed of a mobile e-commerce platform, where a user browses the exposed items, scrolls up and down the feed, clicks an item card to enter the product page, and may further inspect details, add the item to cart, or place an order, spanning the complete exposure-to-purchase funnel.
To turn such sessions into an interactive environment rather than static replay logs, we reconstruct from online logs the complete exposure state of each session, \ie, the full set of items exposed to the user together with their page layouts and interface states.
Consequently, at every step the agent may take any executable action, not merely the one the real user happened to take, and the environment renders the corresponding real interface as visual feedback; this property is what enables genuine multi-turn rollouts for agentic RL.

The closest efforts to ours are STA~\citep{Zhang2025SeeTA} and A/B Agent~\citep{Zhang2026ExploringRS}.
STA replays fixed single-session web logs: once the agent deviates from the logged action at any step, the environment cannot return a genuine next state, so optimization degenerates to per-step prediction over static logs and the exploration required by multi-turn RL becomes infeasible.
On the other hand, A/B Agent instead synthesizes mock interfaces from MovieLens~\cite{harper2015movielens} and Amazon Fashion~\cite{ni-etal-2019-justifying}; such synthetic pages cannot recover the visual environment that real users actually observed, leaving a persistent gap between simulated and real interfaces that undermines the sim-to-real vision.

\begin{figure*}[t]
  \centering
  \includegraphics[width=\textwidth]{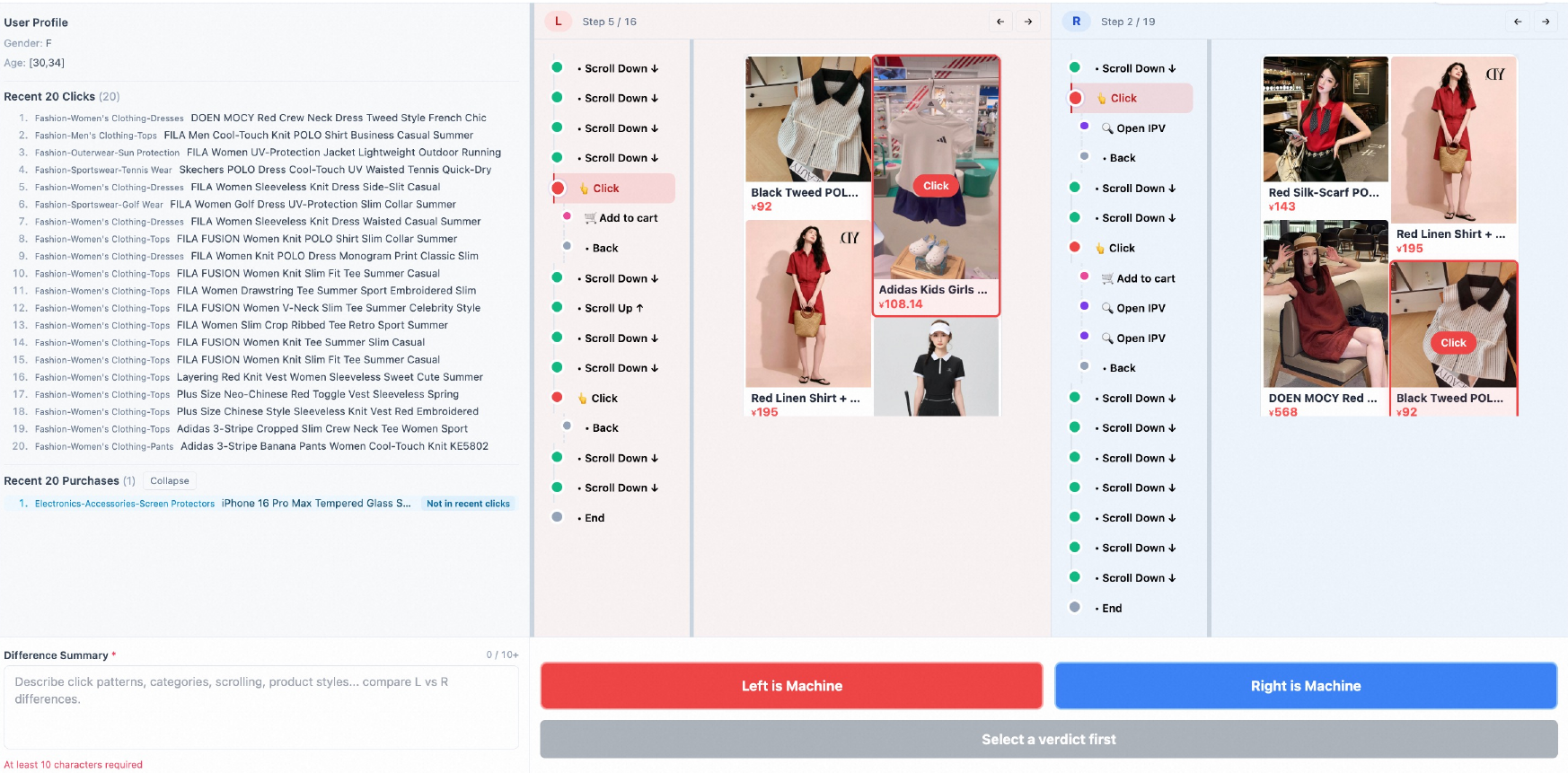}
  \caption{Annotation interface for human evaluation. Annotators compare two anonymized trajectories conditioned on the same user profile, inspect their action timelines and GUI observations, and identify the trajectory judged to be machine-generated.}
  \label{fig:turing_screen}
\end{figure*}

Looking forward, we plan to integrate \ours{} into our production pipeline, where it can serve as (i) a training environment for RL-based recommenders that optimize long-term user engagement and commercial value, (ii) a low-cost offline surrogate for online A/B testing, (iii) a data-augmentation generator for model training, and (iv) a transparent, inspectable testbed for in-depth analysis of user behavior and systematic evaluation of recommender systems.

\section{Experimental Details}
\label{app:exp_details}

\subsection{Baseline Details}
\label{app:baseline_details}
We compare \ours{} against representative user simulation methods across different observation modalities and training strategies.

\paragraph{\underline{\textbf{Text-Based Methods}}}
These methods operate on textual item descriptions, without perceiving visual page observations.
\begin{itemize}[leftmargin=1.2em, itemsep=2pt, topsep=2pt, label=$\diamond$]
    \item \textbf{RecAgent}~\citep{wang2025user} builds LLM-based user agents with profile, memory, and action modules, simulating multi-turn browsing via in-context prompting without any training.
    \item \textbf{Agent4Rec}~\citep{zhang2024generative} prompts an LLM to role-play as a user with a persona derived from historical ratings and reviews, generating page-by-page browsing decisions in a training-free manner.
    \item \textbf{OPeRA}~\citep{wang2025opera} simulates e-commerce behavior from structured textual user and item information in a training-free manner.
    \item \textbf{AlignUSER}~\citep{bougie2026alignuser} fine-tunes an LLM on real user trajectories via imitation learning, additionally introducing a world model to predict environment transitions.
    \item \textbf{Shop-R1}~\citep{zhang2026shopr} applies RL with a hierarchical step-level reward combining action-type matching, sub-action attribute scoring, and a self-certainty signal of the agent's reasoning process.
    \item \textbf{Customer-R1}~\citep{wang2025customer} extends the RL paradigm of Shop-R1 with personalized user profiles and difficulty-aware reward weighting to augment the training signal for low-frequency actions.
\end{itemize}

\paragraph{\underline{\textbf{GUI-Based Methods}}}
These methods share the same visual observation modality as \ours{}, perceiving page-level screenshots.
\begin{itemize}[leftmargin=1.2em, itemsep=2pt, topsep=2pt, label=$\diamond$]
    \item \textbf{A/B Agent}~\citep{Zhang2026ExploringRS} prompts VLMs with visual observations and hand-designed decision and memory modules for A/B testing evaluation, without task-specific fine-tuning.
    \item \textbf{STA} (See-Think-Act)~\citep{Zhang2025SeeTA} is the strongest existing GUI-based baseline, adopting a See-Think-Act pipeline with both IL and RL training; it conditions on the current screenshot together with full action history and heuristically pruned historical HTML observations, and optimizes with step-level rewards.
\end{itemize}

\subsection{Implementation Details}
\label{app:implementation_details}
\paragraph{\underline{\textbf{Training Setup}}}
All experiments use Qwen3.5-2B~\cite{qwen3.5} as the backbone model. Both imitation learning and reinforcement learning adopt full-parameter fine-tuning, while freezing the visual encoder to preserve pretrained visual representations and stabilize multimodal training. We conduct distributed training with Megatron-LM~\cite{shoeybi2019megatron} and use vLLM~\cite{kwon2023efficient} to accelerate rollout inference. For imitation learning, we train for 10 epochs with a learning rate of $2\times10^{-7}$ and a cosine learning-rate schedule. For GRPO, we sample 8 responses per generation, with a total rollout batch size of 16, a learning rate of $2\times10^{-7}$, a KL coefficient of $1\times10^{-3}$, and train for 100 optimization steps. The action weights $w(a)$ in Eq.~\eqref{eq:micro_reward} are set empirically according to the action-frequency distribution in the dataset: click is assigned weight 1, entering a product detail page is assigned weight 5, add-to-cart is assigned weight 5, and purchase is assigned weight 10. We cap both training trajectories and inference trajectories at 20 steps, and set the maximum image input pixel budget to 200{,}704 (\ie, $448\times448$) to balance visual quality and computational cost.

\paragraph{\underline{\textbf{Training Data Construction}}}
We synthesize imitation-learning data from real user trajectories using Qwen3.5-397B-A17B with temperature 0.7. The goal is to warm-start the policy before reinforcement learning with trajectories that follow the reasoning and memory-update format of \ours{}. At each step $t$, given the user profile $u$, the hierarchical memory state $\mathcal{M}_{t-1}$ accumulated over previous steps, the visual observation $o_t$, and the logged next action $a_t^*$, the teacher model performs counterfactual reconstruction of the user's latent state and memory updates:
\begin{equation*}
    z_t, e_t, p_t = \pi_{\text{teacher}}(u, \mathcal{M}_{t-1}, o_t, a_t^*),
\end{equation*}
where $z_t$ denotes the inferred user mindset, and $e_t$ and $p_t$ denote the episodic and preference memory updates, respectively. The memory state is then updated according to the generated memory content and used to synthesize the next step, continuing until the trajectory terminates. Outputs are organized in JSON format to match the structured action and memory schema used by \ours{}. For historical user click and interaction sequences, we truncate the context to at most 20 entries to avoid excessive prompt length.

\begin{table*}[t]
\centering
\caption{Ablation analysis of matching granularity in the micro-level reward. Behavioral fidelity is reported as absolute deviation from real-user statistics; intent metrics are reported as percentages.}
\label{tab:micro_reward_design}
\vspace{-5pt}
\begin{tabular}{lccccc ccccccc}
\toprule
\multirow{2}{*}{\textbf{Matching}} & \multicolumn{5}{c}{\textbf{Behavioral Fidelity} ($\downarrow$)} & \multicolumn{7}{c}{\textbf{Intent Consistency} ($\uparrow$)} \\
\cmidrule(lr){2-6}\cmidrule(lr){7-13}
& $\Delta$ATL & $\Delta$CTR & $\Delta$ACR & $\Delta$CVR & $\Delta$IPVR & P & R & F1 & HR & CP & CR & HCO \\
\midrule
Item & \textbf{1.87} & 3.63 & \textbf{1.65} & 10.66 & 4.22 & 5.46 & 4.45 & 4.90 & 6.51 & 28.81 & 26.14 & 27.41 \\
Category & 2.78 & \textbf{1.30} & 4.06 & \textbf{8.67} & \textbf{3.26} & \textbf{7.41} & \textbf{6.98} & \textbf{7.19} & \textbf{10.45} & \textbf{34.51} & \textbf{30.95} & \textbf{32.64} \\
\bottomrule
\end{tabular}
\end{table*}

The raw action distribution in real browsing trajectories is highly long-tailed, where directly training on the original distribution can cause the policy to collapse to high-frequency navigation actions such as scrolling down. To mitigate this imbalance, inspired by prior work~\cite{conneau2020unsupervised,lample2019cross,arivazhagan2019massively}, we apply \emph{power-law smoothed resampling} over action categories. Let $n_i$ denote the number of samples from action category $i$. The target sampling distribution is defined as:
\begin{equation}
    q_i = \frac{n_i^\alpha}{\sum_j n_j^\alpha},
\end{equation}
where $0 < \alpha < 1$ compresses the frequency gap between head and tail categories. This increases the sampling probability of minority actions while preserving the relative structure of the original distribution, avoiding the overfitting risk of fully balanced resampling. We set $\alpha=0.3$ in our experiments and apply the same resampling strategy to all trainable baselines for fair evaluation.

\subsection{Prompt Template}
\label{app:prompt_template}

For reproducibility, Table~\ref{tab:prompt_template} summarizes the rollout prompt used by \ours{}. We report a condensed template rather than the fully expanded prompt, since page-specific executable actions and visual observations are filled dynamically at each step.

\section{Additional Analysis}
\label{app:additional_analysis}

\subsection{Effect of Matching Granularity in Micro-Level Reward}
\label{app:micro_reward_design}

To examine the matching granularity in the micro-level reward, we compare the default category-matching design with a stricter item-level alternative. Specifically, we replace the category-based score $r(a)$ in Eq.~\eqref{eq:micro_reward} with a strict item-hit score:
\begin{equation}
    r_{\text{item}}(a) = \mathbb{I}(x_a \in \mathcal{I}^*),
\end{equation}
where $x_a$ is the item targeted by action $a$ and $\mathcal{I}^*$ is the reference item set.
This score gives credit only when the simulated action reaches an item that appears in the logged trajectory, whereas category matching allows semantically related items to receive partial credit.

As shown in Table~\ref{tab:micro_reward_design}, category matching provides substantially stronger intent guidance than strict item matching. It improves F1 from 4.90 to 7.19, HR from 6.51 to 10.45, and HCO from 27.41 to 32.64, while also yielding smaller deviations in CTR, CVR, and IPVR. Although strict item matching is closer on ATL and ACR, this advantage does not translate into better intent consistency. These results suggest that exact item overlap is too restrictive as an RL signal for e-commerce simulation: by assigning credit to semantically aligned alternatives, category matching provides denser reward feedback for plausible item-directed decisions.

\subsection{Qualitative Case Study}
\label{app:case_study}

\begin{figure}[t]
  \centering
  \includegraphics[width=\columnwidth]{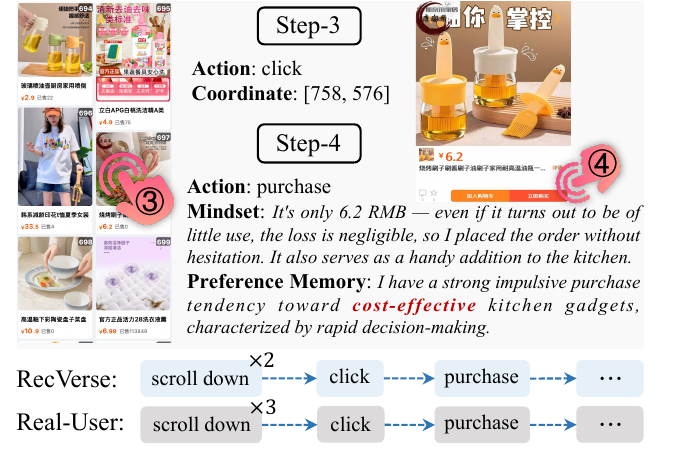}
  \vspace{-15pt}
  \caption{Case study of \ours{} on a real browsing session. The example illustrates aligned action trajectories between \ours{} and the real user, along with generated mindset and preference memory that support the purchase decision.}
  \label{fig:case_study}
  \vspace{-10pt}
\end{figure}

Figure~\ref{fig:case_study} provides a concrete example of how \ours{} simulates a real browsing session. The real user and \ours{} follow a similar high-level trajectory: both first browse the product list, then click a low-price household item, and finally proceed to purchase. Although the exact number of scroll operations differs slightly, the simulated rollout preserves the main behavioral pattern of short exploration followed by a focused purchase decision.

The generated internal states further explain why the action is plausible. The mindset describes the purchase as a low-risk and useful kitchen addition, while Preference Memory abstracts the behavior into an interest in cost-effective kitchen gadgets rather than merely recording the clicked item. This example illustrates the role of hierarchical memory in making simulated behavior both traceable and intent-aware: recent visual context supports the immediate action, and accumulated preference evidence provides continuity for long-term decision-making.

\section{Human Evaluation Interface}
\label{app:human_eval_interface}

Figure~\ref{fig:turing_screen} shows the annotation interface used for human evaluation. Each task presents the annotator with the user's profile and historical interactions, together with two anonymized trajectories displayed side by side. For each trajectory, the interface shows the action sequence, the corresponding GUI observations, and the interacted items highlighted on the page. Annotators are asked to compare the two trajectories, write a brief difference summary, and select the one judged to be machine-generated. This layout provides both profile-level context and step-level visual evidence, while keeping method identities hidden during annotation. We further quantify annotation reliability with Fleiss' $\kappa=0.834$ across annotators and an average pairwise Cohen's $\kappa=0.815$, indicating high labeling consistency among the outsourced annotators.

\end{document}